\documentclass{aa}
\usepackage[labelfont=bf,textfont=footnotesize]{caption}  %% as A&A
\usepackage{etoolbox}
\makeatletter
\patchcmd\@combinedblfloats{\box\@outputbox}{\unvbox\@outputbox}{}{\errmessage{\noexpand patch failed}}
\makeatother
\usepackage[varg]{txfonts}
\usepackage{epsfig,graphicx,natbib,url,twoopt}
\usepackage[varg]{txfonts}
\usepackage[colorlinks=true,urlcolor=blue,citecolor=blue,linkcolor=blue,breaklinks=true]{hyperref}
\usepackage{gensymb}
\usepackage{rotate}

\let\oldhref\href
\renewcommand{\href}[2]{\oldhref{#1}{\hbox{#2}}}

\bibpunct{(}{)}{;}{a}{}{,} %% natbib format for A&A and ApJ
\makeatletter
\newcommandtwoopt{\citeads}[3][][]{\href{http://adsabs.harvard.edu/abs/#3}%
{\def\hyper@linkstart##1##2{}%
\let\hyper@linkend\@empty\citealp[#1][#2]{#3}}}
\newcommandtwoopt{\citepads}[3][][]{\href{http://adsabs.harvard.edu/abs/#3}%
{\def\hyper@linkstart##1##2{}%
\let\hyper@linkend\@empty\citep[#1][#2]{#3}}}
\newcommandtwoopt{\citetads}[3][][]{\href{http://adsabs.harvard.edu/abs/#3}%
{\def\hyper@linkstart##1##2{}%
\let\hyper@linkend\@empty\citet[#1][#2]{#3}}}
\newcommandtwoopt{\citeyearads}[3][][]%
{\href{http://adsabs.harvard.edu/abs/#3}
{\def\hyper@linkstart##1##2{}%
\let\hyper@linkend\@empty\citeyear[#1][#2]{#3}}}
\makeatother

\newcommand{\bq}{\begin{equation}}
\newcommand{\eq}{\end{equation}}
\newcommand{\beaq}{\begin{eqnarray*}}
\newcommand{\eeaq}{\end{eqnarray*}}

\newcommand{\halpha}{\mbox{H\hspace{0.1ex}$\alpha$}}

\def\kms{\hbox{km$\;$s$^{-1}$}}

\newcommand{\efft}{\ion{Ca}{ii}~8542~\AA}
\newcommand{\caii}{\ion{Ca}{ii}}
\newcommand{\mgii}{\ion{Mg}{ii}}
\newcommand{\siiv}{\ion{Si}{iv}}
\newcommand{\oiv}{\ion{O}{iv}~1401~\AA}  
\newcommand{\cii}{\ion{C}{ii}} 
\newcommand{\feii}{\ion{Fe}{ii}} 
\newcommand{\cahline}{\ion{Ca}{ii}~H line}
\usepackage{epstopdf}
\usepackage{epsfig}
\usepackage{threeparttable}
\begin{document}  

\title{A multi-diagnostic spectral analysis of penumbral microjets}

\author{Ainar Drews\inst{1,2}
\and
Luc Rouppe van der Voort\inst{1,2}
}
\authorrunning{L. Rouppe van der Voort \& A. Drews}

\institute{Institute of Theoretical Astrophysics,
  University of Oslo, %
  P.O. Box 1029 Blindern, N-0315 Oslo, Norway
\and
 Rosseland Centre for Solar Physics,
  University of Oslo, %
  P.O. Box 1029 Blindern, N-0315 Oslo, Norway}

\date{\today}

\abstract
 {
 Penumbral microjets (PMJs) are short-lived, jet-like objects found in the penumbra of sunspots. They were first discovered in chromospheric lines and have later also been shown to exhibit signals in transition region (TR) lines. Their origin and manner of evolution is not yet settled.
 }
 {%Aims
 We perform a comprehensive analysis of PMJs through the use of spectral diagnostics that span from photospheric to TR temperatures to constrain PMJ properties.
 }
 {%Methods
  We employed high-spatial-resolution Swedish 1-m Solar Telescope observations in the \efft\ and \halpha\ lines, IRIS slit-jaw images, and IRIS spectral observations in the \mgii\ h \& k lines, the \mgii\ $2798.75$ \AA\ \& $2798.82$ \AA\ triplet blend, the \cii\ 1334 \AA\ \& 1335 \AA\ lines, and the \siiv\ 1394 \AA\ \& 1403 \AA\ lines. We derived a wide range of spectral diagnostics from these and investigated other secondary phenomena associated with PMJs.
 }
 {%Results
 We find that PMJs exhibit varying degrees of signal in all of our studied spectral lines. We find low or negligible Doppler velocities and velocity gradients throughout our diagnostics and all layers of the solar atmosphere associated with these. Dark features in the inner wings of \halpha\ and \efft\  imply that PMJs form along pre-existing fibril structures. We find evidence for upper photospheric heating in a subset of PMJs through emission in the wings of the \mgii\ triplet lines. There is little evidence for ubiquitous twisting motion in PMJs. There is no marked difference in onset-times for PMJ brightenings in different spectral lines.
 }
 {%Conclusions
 PMJs most likely exhibit only very modest mass-motions, contrary to earlier suggestions. We posit that PMJs form at upper photospheric or chromospheric heights at pre-existing fibril structures.
 }

\maketitle

\section{Introduction \label{sec:introduction}} 

Penumbral microjets (PMJs) are observed in the penumbrae of sunspots, and they were first discovered in the 
chromospheric \cahline\
\citepads{2007Sci...318.1594K}. 
Time-sequences produced by the Hinode satellite's 3 \AA\ wide \cahline\ imaging filter revealed short-lived 
jet-like objects which were most noticeable in time-difference images. In these original observations, PMJs show 
a relative brightening of 10\%\ - 20\%\ compared to their penumbral surroundings, have lengths of 1000 km - 4000 km and 
widths of about 400 km, typical lifetimes of up to 1 minute, and apparent rise-velocities of 100 \kms\ 
\citepads{2007Sci...318.1594K}.
In 
\cite{2007Sci...318.1594K}, 
the authors speculated that the fast apparent velocities of PMJs could be explained by either a true mass motion caused by a reconnection outflow that exceeds acoustic velocities or alternatively by the evolution of a thermal conduction front that lacks 
significant mass motions.
Through inversions and analyses of PMJ observations, 
\cite{2019ApJ...870...88E}
posit that PMJs are the result of a propagating perturbation front that originates in the deep photosphere and dissipates energy within the PMJ in a process akin to what has been suggested for spicules in recent times 
\citep{2017ApJ...849L...7D}. 
This lends support to the thermal conduction front scenario of
\cite{2007Sci...318.1594K}.  

A large automated sampling of PMJs 
\citepads{2017A&A...602A..80D} 
using highly spatially, temporally, and spectrally resolved \efft\ observations from the 
Swedish 1-m Solar Telescope (SST) at La Palma gives average PMJ lengths, widths, and lifetimes of 640 km, 210 km, and 90 s (with an 8 minute cut-off), 
respectively, on the same order of magnitude as previous values. PMJs are thought to be chromospheric in origin 
\citep{2007Sci...318.1594K, 2010A&A...524A..21J}, 
but they have been shown to have transition region (TR) responses 
\citep{2015ApJ...811L..33V, 2016ApJ...816...92T, Katsukawa2018}  
 and are visible in the \ion{Mg}{ii} k, \ion{C}{ii}, and \ion{Si}{IV} slit-jaw images observed by the Interface Region Imaging Spectrograph 
satellite
\citepads[IRIS, ][]{2014SoPh..289.2733D}. 
These observations suggest heating to TR temperatures as the studied PMJs show emission 
in the \ion{C}{ii} and \ion{Si}{IV} lines towards their tops. 

TR bright dots are bright features that are predominantly found in the penumbrae of sunspots 
\citepads{2014ApJ...790L..29T,2014AGUFMSH51C4182A}, 
and they are visible in TR channels, such as the IRIS SJI $1400$ \AA\ and $1330$ \AA\ channels. They are usually somewhat elongated and have sizes on the scale of a few hundred kilometres, and upon discovery they were already speculated to be linked to PMJs.
\cite{2017ApJ...835L..19S} 
posit more specifically that PMJs may in fact originate at TR heights in the form of TR bright dots and show chromospheric signatures only after their TR counterparts, and they report observations to this effect.  
However, the most popular proposed mechanism for the creation of PMJs remains magnetic reconnection in the photospheric penumbra, as initially 
suggested upon their discovery 
\citepads{2007Sci...318.1594K}. 
The magnetic reconnection scenario at photospheric or lower chromospheric heights is supported by the measurement of apparent inclinations of 
PMJs to surrounding penumbral filaments 
\citepads{2007Sci...318.1594K}  
 and magnetic fields 
\citepads{2008A&A...488L..33J}. 
Some small photospheric downflow patches are also observed in conjunction with some PMJs
\citepads{2010A&A...524A..20K}, 
which may further strengthen the assumption of photospheric magnetic reconnection. 
There is also numerical work that suggests the plausibility of the photospheric magnetic reconnection scenario   
\citepads{2012ApJ...761...87N}, showing that reconnection can indeed give rise to jet-like phenomena that travel in the approximate 
direction of the surrounding penumbral fibrils and later reach observed apparent velocities 
as a result of moving from the dense photosphere to the less dense chromosphere by producing a shock.

A distinct PMJ profile in the \efft\ line was first presented by 
\cite{2013ApJ...779..143R}, 
displaying peaks in the inner line wings with a blue-over-red asymmetry. 
The average \efft\ PMJ line profile of 
\citet{2017A&A...602A..80D} 
did not reveal any significant Doppler shift, nor a viewing-angle correlation between the 
line offset of the blue or red peaks in the distinct line profile.

Large penumbral jets (LPJs) were first described in 
\cite{2016ApJ...816...92T} 
as larger-than-average and more energetic PMJs that can be found at the edge of penumbrae. 
\cite{2018ApJ...869..147T}
present evidence that LPJs may exhibit twisting motions along their long axis. This was accomplished through the analysis of spectral profiles observed using IRIS in the \mgii\ k line. At least a subset of the objects studied in this and other works that are termed PMJs may instead have been termed LPJs by Tiwari and collaborators.
As such, the potential twisting motion of LPJs may also be expected of regular PMJs, and more so of large PMJs at the edge of penumbrae, which are in essence defined to be LPJs in the cited works.   

The time evolution of PMJs has only been studied observationally in broad terms, especially when comparing their signatures in different wavelengths and thus solar temperatures through time. Earlier works have focused on lifetimes and apparent velocities of PMJs, usually in specific chromospheric lines. In 
\cite{2019A&A...626A..62R}, 
we argue for a revised view of PMJ temporal evolution in light of highly temporally resolved SST observations of PMJs in the \cahline. PMJs only show modest, true apparent velocities and appear to light up across a significant fraction of their length along existing fibrils in the \cahline. This happens on the timescales of the cadence of the observations of about 1 s. This is hypothesised to be due to a heating front moving through the PMJs, rather than hot material moving upwards in a true mass motion. 
This interpretation is congruent with the proposed scenario mentioned earlier, which was first put forth by 
\cite{2007Sci...318.1594K}, regarding 
an evolving thermal conduction front, which was 
strengthened by the findings of 
\cite{2019ApJ...870...88E}. 
To definitively differentiate  between a scenario involving true high-velocity mass motions or that of an evolving thermal conduction front, we performed a wide range of Doppler-shift measurements, as first suggested in \cite{2007Sci...318.1594K}. 

Here, we expand on the investigation of the TR response of PMJs as performed in \cite{2015ApJ...811L..33V} by analysing co-observations from 
the SST and IRIS, covering the solar atmosphere 
from the photosphere to the TR. We sampled the photosphere and chromosphere of a fully formed sunspot with detailed 
line scans of \efft\ and \halpha\ with the SST and sampled the upper chromosphere and TR 
with slit-jaw images and spectra of \ion{Mg}{ii}, \ion{C}{ii}, and \ion{Si}{IV} from IRIS. We assembled a set of 77 PMJs, which were co-observed with both the SST and IRIS. The PMJs' IRIS signatures were sampled at IRIS' spectrograph slit-positions and at all intermediate pixels covered by the SST in the \efft\ and \halpha\ lines. 

The above enabled us to acquire a wide range of spectral diagnostics at different locations along the PMJs, so that we can analyse spectral features at different nominal heights and throughout the PMJs' evolution, from formation to dissipation and heights from the photosphere to the TR. We also study possible twisting motions of PMJs utilising the \mgii\ line and describe concurrent PMJ features in the inner wings of the \halpha\ and \efft\ lines. Lastly, we describe the appearance of the various PMJ signals in the different spectral channels through time.

%======================================= fig 1
\begin{figure}[t]
\centering
\includegraphics[width=\columnwidth]{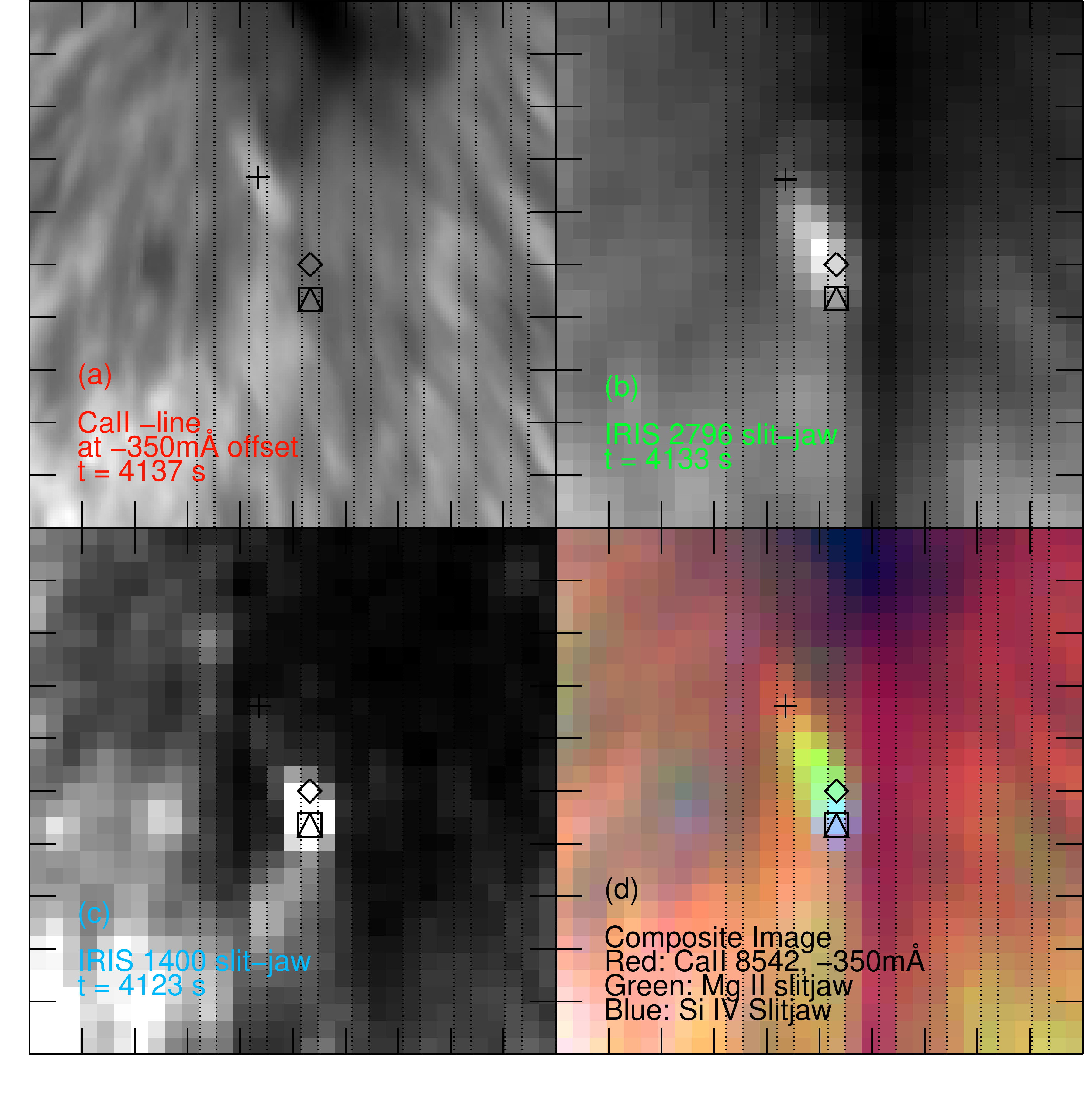} 
\caption{\label{fig:selection} Example PMJ from dataset B. Observations of (a) the SST \efft\ line at an offset of $-350$~m\AA, (b) the IRIS SJI 2796, (c) the IRIS SJI 1400, and (d) the composite RGB image created from the three preceding images are shown. Tick-mark spacing is 1''. The images were cropped  with the PMJ at their approximate centres. Symbols mark the sampling positions for analysis of spectral line profiles. Cross: \efft, diamond: the \mgii\ k, h, and triplet lines, square: \cii\ 1334 \AA\ and 1335 \AA\ lines, triangle: \siiv\ 1394 \AA\ and 1403 \AA\ lines. The slit positions (of one-pixel width) of IRIS are denoted in each panel by black-dotted lines. Tick marks are spaced 1'' apart.}
\end{figure}
%=======================================

\section{Observations \label{sec:observations}}  

We observed NOAA Active Region AR12533 on April 29 and 30, 2016. The field of view for SST and IRIS was aimed at the centre of the oval sunspot with a fully formed penumbra. On April 29, the AR was at heliocentric coordinates (X,Y) = (623'', 19'') ($\mu = \cos \theta = 0.75$, with $\theta$ the observing angle). On April 30, it was observed again, but then at heliocentric coordinates (X, Y) = (774'', 4'') ($\mu = \cos \theta = 0.57$). The SST seeing was good for the duration of the co-pointing on both days. 
The data quality was further improved by the adaptive optics system 
\citepads{2003SPIE.4853..341S}
and subsequent image reconstruction with the multi-object multi-frame blind deconvolution method 
\citepads[MOMFBD, ]{2005SoPh..228..191V}.
The observing procedure and programmes used for both instruments were identical for both sets of observations. We used the CRisp Imaging SpectroPolarimeter  
\citepads[CRISP, ][]{2008ApJ...689L..69S} 
instrument to acquire spectrally resolved data in the \efft\ and \halpha\ spectral lines. Six exposures were acquired per liquid crystal state and line position, and the \efft\ line scan was completed in $\sim16$~s, and with the \halpha\ scan included, the effective cadence of the time sequence was $\sim20$~s. The spectropolarimetric \efft\ data sampled the spectral line at 21 line positions, with a dense (70 m\AA\  steps) sampling in the line core and an increasingly coarser sampling in the wings out to $\pm1.75$~\AA. 
The \halpha\ line was sampled at 15 positions, with 200 m\AA\ steps around the line core and the last two sampling positions at $\pm1500$~m\AA.

The CRISP data were processed with the CRISPRED 
reduction pipeline
\citepads{2015A&A...573A..40D}. 

IRIS ran a so-called medium sparse eight-step raster observing programme (OBSID 3620106129) with a 60\arcsec\ long and 0\farcs33 wide spectrograph slit covering a 7\arcsec\ wide area with eight slit positions separated by 1\arcsec. The exposure time was 4~s and the spectrograms were 2$\times$ binned both in the spatial (0\farcs33 per pixel) and spectral domain (26 m\AA\ per pixel for the FUV spectra and 51 m\AA\ per pixel for the NUV). The raster cadence was $\sim40$~s. Slit-jaw images (SJI) were recorded in the SJI 1400~\AA, 1330~\AA, and 2796~\AA\ channels at a cadence of $\sim20$~s, and in the SJI 2832~\AA\ channel at a 122~s cadence. All slit-jaw images were spatially binned to 0\farcs33 per pixel.

Co-pointing proved to be successful, with both instruments imaging the whole sunspot for the two time series at 09:42 -11:13 UT and 09:08 -10:38 UT on April 29 and 30, 2016 respectively. The SST and IRIS observations were co-aligned through cross-correlation between the \efft\ wing and the SJI 2832 Mg h wing channel.

Going forward, when referring to the observations from April 29, April 30 and the values derived thereof, we use the short-hand terms dataset A and dataset B, respectively. 

\section{Methods \label{sec:methods}}

To investigate the spectral signatures of PMJs in the wavelength passbands available from the 
SST and IRIS observations, we identified examples of PMJs using the  CRisp SPectral EXplorer 
\citepads[CRISPEX,]{2012ApJ...750...22V}.
This allowed us to view the observations simultaneously in the SST \efft\ line and the IRIS slit-jaw 
images, allowing for easier initial identification of sample events crossing the IRIS raster-positions. For the final identification and selection of suitable PMJs that crossed the IRIS spectrograph raster slits, we employed a visual examination of side-by-side images in the SST and SJI IRIS observations together with composite images of them. 

PMJs are the brightest at an offset of $-350$~m\AA\  in the \efft\ line \citepads{2017A&A...602A..80D}, 
corresponding to the typical blue peak position in this line. We 
created RGB images from  \efft\ $-350$~m\AA\ images (Red), SJI 2796 images (Green), and SJI 1400 images (Blue). IRIS slitjaw images and 
images in the \efft\ blue wing can be employed to highlight the progression of PMJs through temperatures and corresponding heights that usually correspond to these channels. The inner wings of the CRISP \efft\ line are usually associated with the chromosphere, the IRIS SJI 2796 with the upper cromosphere, and the IRIS SJI 1400 with the TR.

The RGB images were adjusted for contrast and 
intensity ranges in order to aid in qualitatively identifying PMJs. In these images, PMJs often exhibit a distinct `rainbow' signature. This arises from 
the fact that PMJs typically exhibit spatially offset enhancements in these channels through the different atmospheric heights. This signature is typically orientated along the same direction as the jet-like structure of the PMJ. This behaviour was first described in  
\cite{2015ApJ...811L..33V}. 
An example of a PMJ in the different mentioned channels and the resultant RGB composite image that displays a rainbow signature is given in Fig. \ref{fig:selection}. This and other similar figures aided in the further identification of PMJs. 

There exists a bias towards the detection of PMJs that have a typical line-profile shape in the \efft\ line (as is described and documented in
\citealt{2017A&A...602A..80D}), 
as these were assumed to be a general identifying feature of PMJs. As such, PMJs are assumed to exhibit a signal and typical \efft\ line profiles in the vast majority of cases, and we further assume that PMJs that do not exhibit such typical profiles are either rare or cannot truly be classified as PMJs, as these profiles are a firmly established feature of PMJs. In some detected cases, \efft\ profiles are subdued or absent, but the corresponding PMJs were still included as long as they otherwise appeared typical, visually speaking.

We justify the above in that we canonically consider PMJs to be primarily chromospheric features, as they would otherwise be indistinguishable from bright dots, for example, and a delineating definition is therefore necessary. 
In \citealt{2017ApJ...835L..19S}, 
a causal connection between TR bright dots and PMJs is proposed. Whether or not bright dots may cause or be linked to PMJs should have no bearing on whether PMJs are canonically required to exhibit a signal in the chromosphere, since the lack of signal in the chromosphere for a `PMJ' would imply that it is a bright dot instead.   

Cross-examining and identifying potential PMJs in the different channels, but considering the known and well-described PMJ signatures in 
the \efft\ line, then led to the final selection of PMJs used in further analyses. Selected PMJs were necessarily restricted to those that intersected IRIS slit-positions, so that we could study their spectra.  
The spectral signatures of these events were investigated in a wide range of lines. With CRISP, we obtained spectral profiles in the \efft\ and \halpha\ lines. With IRIS, we obtained spectra of the \mgii\ h \& k and the wavelength region between them, which encloses two of the \mgii\ triplet lines. Further, we obtained spectra of \cii\ 1334 \AA\ and 1335 \AA\ as well as \siiv\ 1394 \AA\ and 1403 \AA. Specific pixel-positions along the raster-slit positions for the IRIS lines and at suitable pixel positions for the CRISP \efft\ line were selected; the aim being to find the positions with the greatest observed response for each line at the approximate time of maximum intensity throughout all channels for all selected PMJs. Each pixel position in each individual channel was selected at a time that was closest in time to the other different sample positions in the other channels, subject to the limitations due to the differences in cadences and exposure times. 

Each PMJ was sampled at four different spatial sampling positions, corresponding to the enumerated spectral lines and PMJ parts below. The PMJ parts are described in reference to when they were viewed in all channels (or their composite RGB images). The sampling positions corresponding to different spectral lines and distinct PMJ parts are the following:
\begin{enumerate}
\item the blue wing of the \efft\ line, corresponding to the chromospheric PMJ footpoint;
\item the \mgii\ h \& k lines (and the two \mgii\ triplet lines and an adjacent \feii\ line), corresponding to the mid-point of the PMJ; 
\item the \cii\ 1334 \AA\ and 1335 \AA\  lines, corresponding to the tail-end of the PMJ; and
\item the \siiv\ 1394 \AA\ and 1403 \AA\ lines, also corresponding to the tail-end of the PMJ.
\end{enumerate}

As mentioned, all of these positions were chosen to maximise the overall spectral line enhancement in each of the relevant wavelength regions and spectral lines. As such, for the selected positions in the IRIS channels, these do not need nor do they frequently correspond to the brightest pixel positions in corresponding IRIS slitjaw images. This is not only due to the positions being limited to the spectral line-raster positions, but also because the slitjaw images correspond to the wide-band wavelength regions of the relevant channels, and finally because the images are also offset in time to the raster samplings (see Sect. \ref{sec:observations}).

We also obtained spectral line profiles from two additional sampling positions by utilising our CRISP observations. We sampled one of these in \efft\ and 
one in \halpha. These sampling positions correspond to dark features in the wings of these two lines that can be found adjacent to the main bright feature identifying PMJs in \efft. All spectral profiles for both \efft\ and \halpha\ were drawn from CRISP observations that were downsampled to match the lower pixel scale of the IRIS slitjaw images in order to select appropriate positions concurrent with the slitjaw-selected IRIS-channel positions.    

After initial identifications, the spectral signatures of PMJs through time at different positions were investigated. 
Our spectral analysis includes (among others) the detection of the wavelength positions of peaks and minima in those lines where relevant, and this was followed by the computation of inferred values, such as peak separations, ratios between the intensities of peaks, and others. 

The different diagnostics are described in Sect. \ref{sec:results} in the various subsections in which they are presented. The temporal evolution of PMJs was studied in a predominantly qualitative fashion. The approximately 40 second cadence of each individual slit-position of the IRIS raster spectrograph makes a detailed 
study of behaviour in time tenuous at best, but it does allow for generalised statements about the behaviour across different channels and therefore associated heights in the solar atmosphere.

\section{Results \label{sec:results}}

This section is organised as follows. We begin by presenting an overview of the detected PMJs in both of our sets of observations in Sect. \ref{sec:results_overview}. Here we point to the general appearance of the sampled PMJs and in which diagnostic channels and with what instruments they are detectable, and we present their general appearance in different spectral lines. We also present the first-order signals of the PMJs in all our diagnostics, meaning their visual appearance in images and their appearance in individual spectral lines as detected by our different instruments. We also present average spectral line profiles for all of our PMJs. Here, we also present size estimates for our PMJ events. 

In Sect. \ref{sec:results_8542} through \ref{sec:results_oiv}, we then present the specific behaviour in particular spectral lines and wavelengths, as well as the second order diagnostic values inferred from signals in specific spectral lines. These second-order values consist of line profile peak positions, peak separations, line-intensity ratios, and more. We present them where relevant for all of our detected PMJs. Here we also present inferred physical conditions in the solar atmosphere at the site of our studied PMJs as estimated from the second-order diagnostic values.

The sections pertaining to the spectral line analysis of the IRIS line profiles lean heavily on the work presented in the series of papers on \emph{The Formation of IRIS Diagnostics}, 
and more specifically the papers dealing with the formation and the analysis of the \mgii\ h, k, the nearby triplet lines \citep{2013ApJ...772...89L,2013ApJ...772...90L,2015ApJ...806...14P}, and the \cii\ 1334 \AA\ and 1335 \AA\ lines \citep{2015ApJ...811...80R,2015ApJ...811...81R}.    

After spectral analysis of our PMJs, we performed three more general analyses, which are presented separately. In Sect. \ref{sec:results_halpha} we present our investigation of dark features observed in the inner line-wings of the \halpha\ line linked to PMJs (first observed in
\cite{2019ApJ...876...47B}), 
which we link to similar darkenings in the inner line wings of the 
\efft\ line. We further link the darkenings at these wavelengths to the brightenings observed in the \mgii\ line pair.
In Sect. \ref{sec:results_mgii_doppler} we investigate the possibility of twisting in PMJs utilising the \mgii\ h \& k lines using Doppler maps and bisectors.
Finally, in Sect. \ref{sec:time_evolution} we perform a qualitative analysis of the temporal behaviour of our studied PMJs through time across different spectral lines.

\subsection{Overview of detected PMJs}\label{sec:results_overview}

%======================================= fig 2
\begin{figure}[!htb]
\centering
\includegraphics[width=7.5cm]{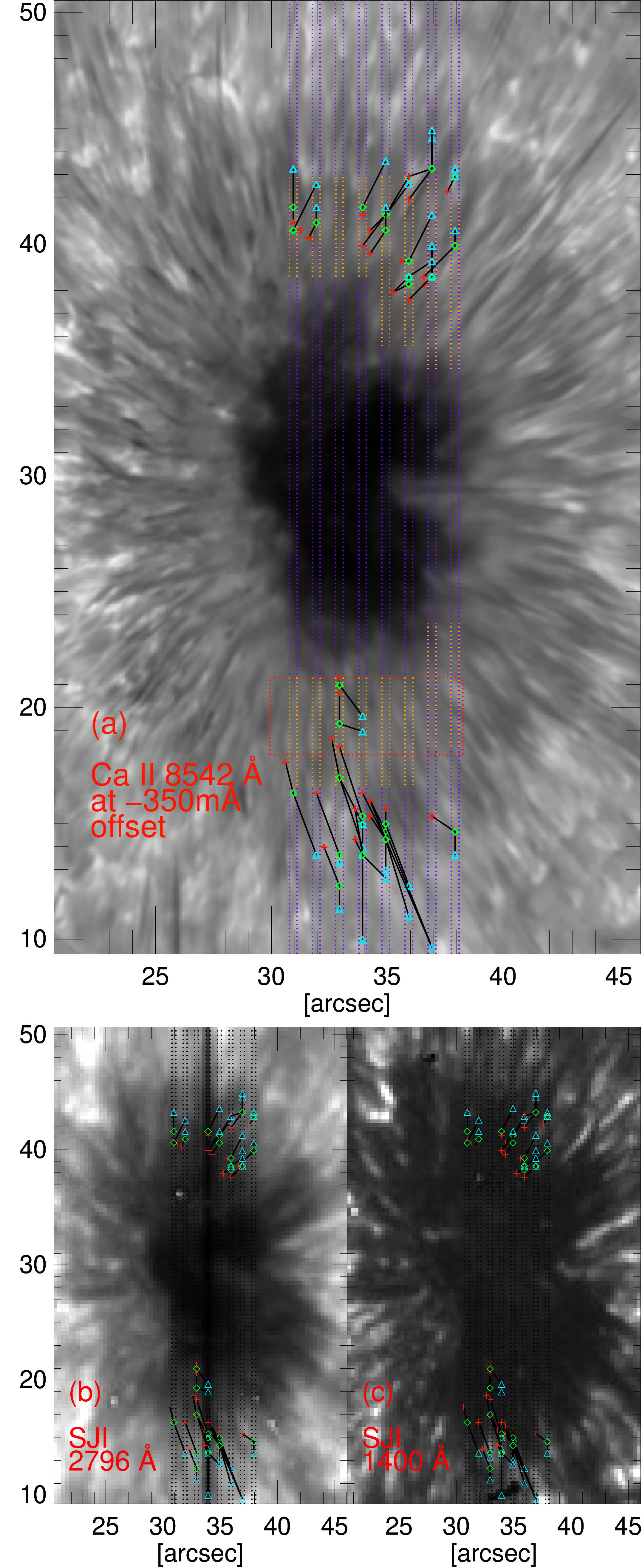} 
\caption{Cropped field of view of AR12533 on April 29, 2016 (dataset A), showing sampling positions for all PMJs studied on this date. Observations of 
(top) the SST \efft\ line at an offset of $-350$~m\AA, (bottom left) the IRIS SJI \mgii\ k line image, and (bottom right) the IRIS SJI \siiv\ line image are shown.
Symbols mark a subset of all the sampling positions for analysis of spectral line profiles. Red cross: \efft; green diamond: the \mgii\ k, h, and triplet lines; and blue triangle: \siiv\ 1394 \AA\ and 1403 \AA\ lines. The raster-slit positions (of one-pixel width) of the IRIS slit-spectrograph are indicated and enclosed in each panel with purple-dashed lines. The mean spectral profiles for the penumbra in the IRIS lines were calculated in the yellow dash marked raster-slit areas. Penumbra mean spectral profiles in the CRISP lines were computed within the red dashed box.\label{fig:all_pmjs29} }
\end{figure}
%======================================
%======================================= fig 3
\begin{figure}[!htb]
\centering
\includegraphics[width=7.5cm]{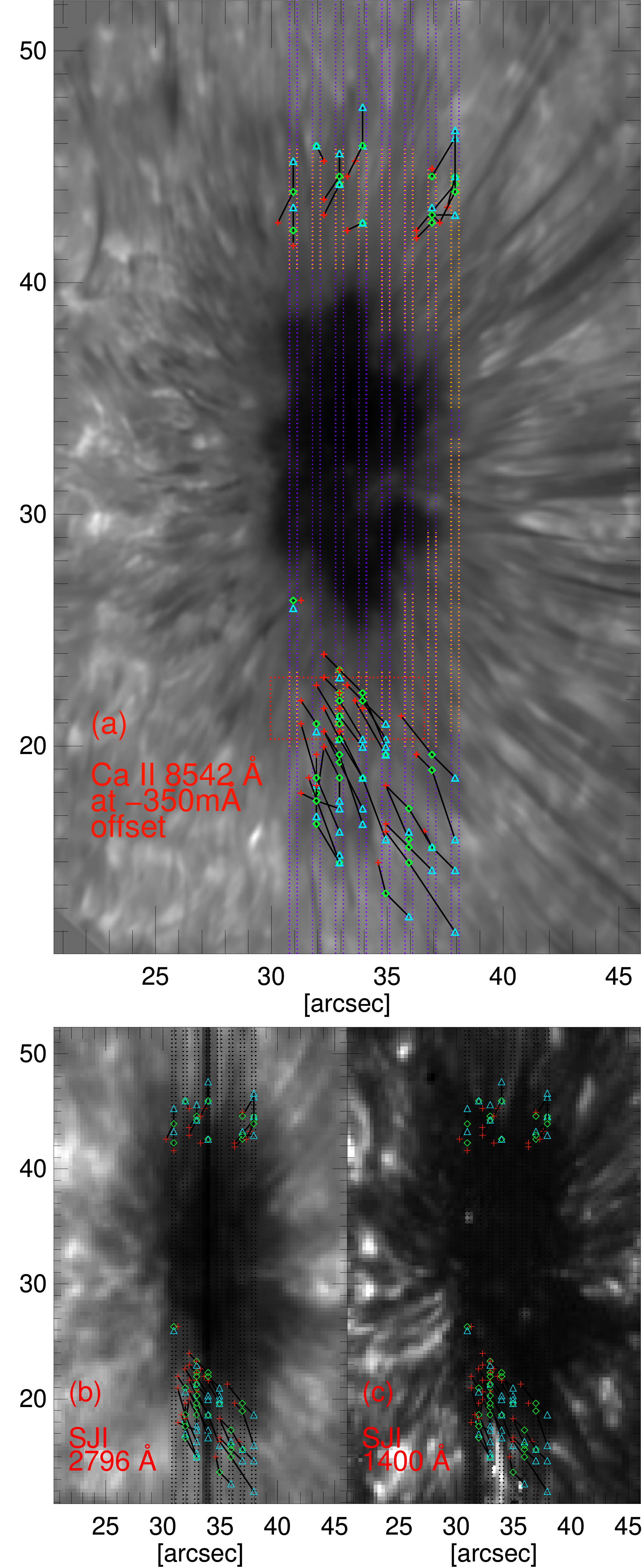} 
\caption{Cropped field of view of AR12533 on April 30, 2016 (dataset B), showing positions for all PMJs studied on this date. The layout is identical to that of Fig. 
\ref{fig:all_pmjs29}, see its caption for details. \label{fig:all_pmjs30} }
\end{figure}
%=======================================

A total of 77 PMJs were detected and found suitable for investigation. Dataset A consists of 33 PMJs, and dataset B is made up of 44 PMJs.
The locations of all detected PMJs and their primary sampling positions in four of our studied channels are shown in Figs. \ref{fig:all_pmjs29} and \ref{fig:all_pmjs30} for datasets A and B, respectively. Only the sampling positions for the \efft\ line, the shared position for the \mgii\ lines, and the position for the \siiv\ lines are shown explicitly for clarity. These positions are representative of the extent of the main body of the investigated PMJs through the different image-channels as these positions usually lie at the foot, midpoint, and terminus of any given PMJ. Also the regions for which we computed the mean spectral profiles of the penumbra for different channels are shown. 
We find PMJs in both the upper and lower portions of the sunspot where these regions are covered by the IRIS raster.
For both datasets, we can see a loose trend of PMJs clustering in so-called hot spots, which is a trend previously described by
\cite{2016ApJ...816...92T}, 
\cite{2017A&A...602A..80D}, 
and
\cite{2019ApJ...870...88E}. 
As can be seen in Figs. \ref{fig:all_pmjs29} and \ref{fig:all_pmjs30}, for both dates of observations, there are two clusters of PMJs to the left and right in both the upper and lower parts of the sunspot. The trend is more clearly visible for dataset B, which is likely due to the higher number of studied PMJs in these observations. The different clusters seem to persist in the sunspot between the two different dates, at least to some degree. This reaffirms the clustered appearance of PMJs in such hot-spots. However, there is a necessary selection bias towards PMJs situated in the region covered by the IRIS raster, as only those PMJs that were caught along most of their length by the raster were selected for investigation. Especially for the lower portion of the sunspot for both dates, this creates a preference for PMJs on the left side of the raster space, as the fibril direction trends from left to right in this area. PMJs typically align roughly with the underlying fibrilar structure, and as such the heads of PMJs lie to the right in the present observations, making the selection of PMJs with footpoints on the left side of the raster region more likely. 

\subsubsection{Videos of PMJ diagnostics}
\label{sec:pmj_video_intro}

In order to facilitate a holistic overview of individual PMJs using all major first-order diagnostics and their temporal evolution, we created two sets of videos for each dataset that show individual PMJs. Two videos show each PMJ in both datasets at their peak brightness, while the second set shows all PMJs through time for both datasets.

The videos showing all PMJs at only peak brightness for dataset A and B are available in \href{https://www.mn.uio.no/astro/personer/vit/rouppe/movies/drews_2020_appa_video1.mp4}{Video 1} and 
\href{https://www.mn.uio.no/astro/personer/vit/rouppe/movies/drews_2020_appa_video2.mp4}{Video 2,}
respectively.
The videos detailing the temporal evolution of PMJs in dataset A and B are available in \href{https://www.mn.uio.no/astro/personer/vit/rouppe/movies/drews_2020_appa_video3.mp4}{Video 3} and 
\href{https://www.mn.uio.no/astro/personer/vit/rouppe/movies/drews_2020_appa_video4.mp4}{Video 4,}
respectively.

A more detailed description of these videos and their layout is given in Appendix \ref{sec:appendix_videos}. We strongly encourage readers to view at least videos 1 and 2 for an overview of all PMJs. Videos 3 and 4 are useful in order to gain an overview of the temporal evolution of specific PMJs. We also particularly suggest inspecting the example PMJs mentioned in the text.

\subsubsection{Visual appearance of PMJs}
%======================================= fig 4

\begin{figure}[!h]
\centering
\includegraphics[width = 8.2cm]{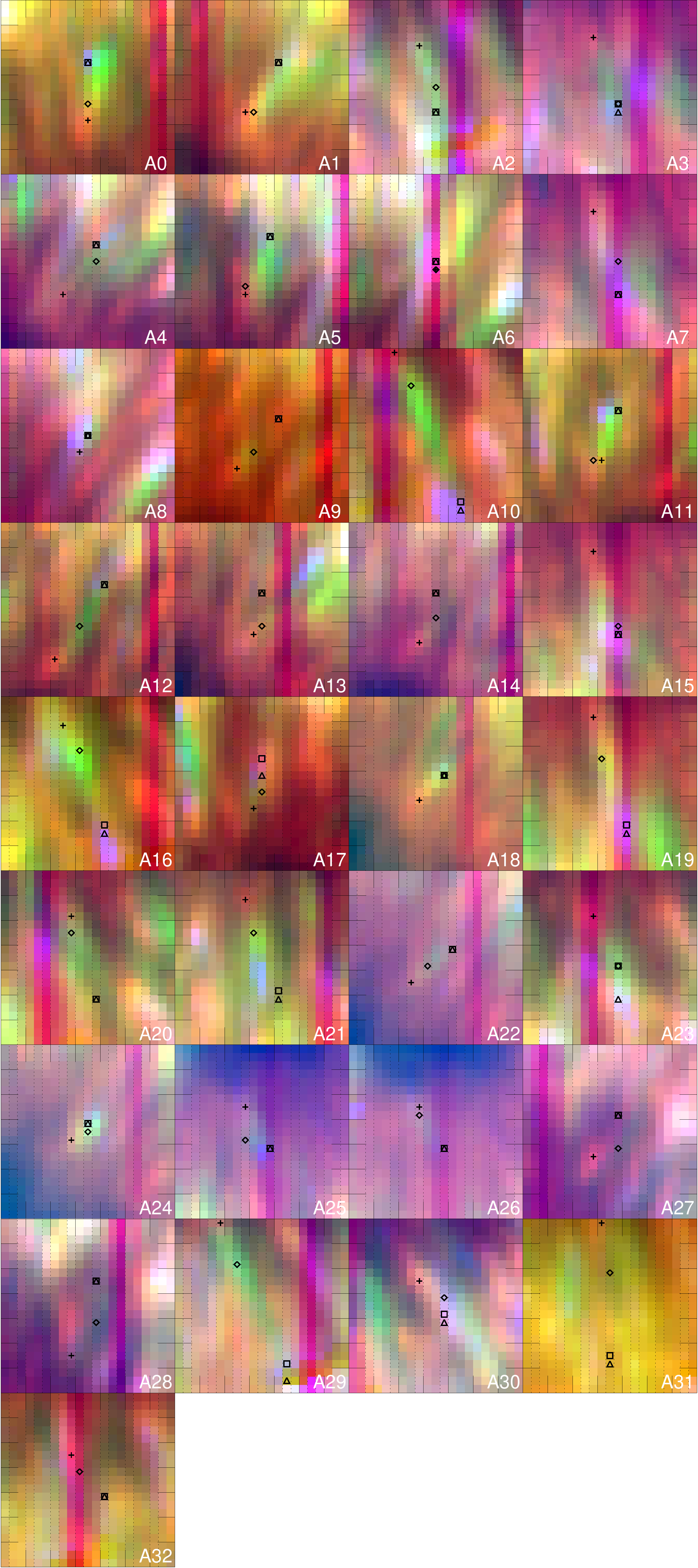} 
\caption{\label{fig:rainbows_29} All PMJs from dataset A. Each panel highlights one 
PMJ at approximately peak intensity. The field of view for each panel is 7 arcseconds along the x- and y-axis and is centred on each PMJ displayed. The panels display colour composite images produced from \efft\ at an offset of $-350$~m\AA\ and the IRIS SJI images in the \mgii\ and \siiv\ lines, which are coded to the colours red, green, and blue, respectively. Sampling positions for the analysis of the different spectral lines are indicated. Plus-sign: \efft; diamond: the \mgii\ k, h, and triplet lines; square: \cii\ 1334 \AA\ and 1335 \AA; and triangle: \siiv\ 1394 \AA\ and 1403 \AA. The raster-slit positions (of one-pixel width) of the IRIS slit-spectrograph are indicated and enclosed in each panel with black-dotted lines.}
\end{figure}
%=======================================

%======================================= fig 5
\begin{figure}[!h]
\centering
\includegraphics[width = 8.2cm]{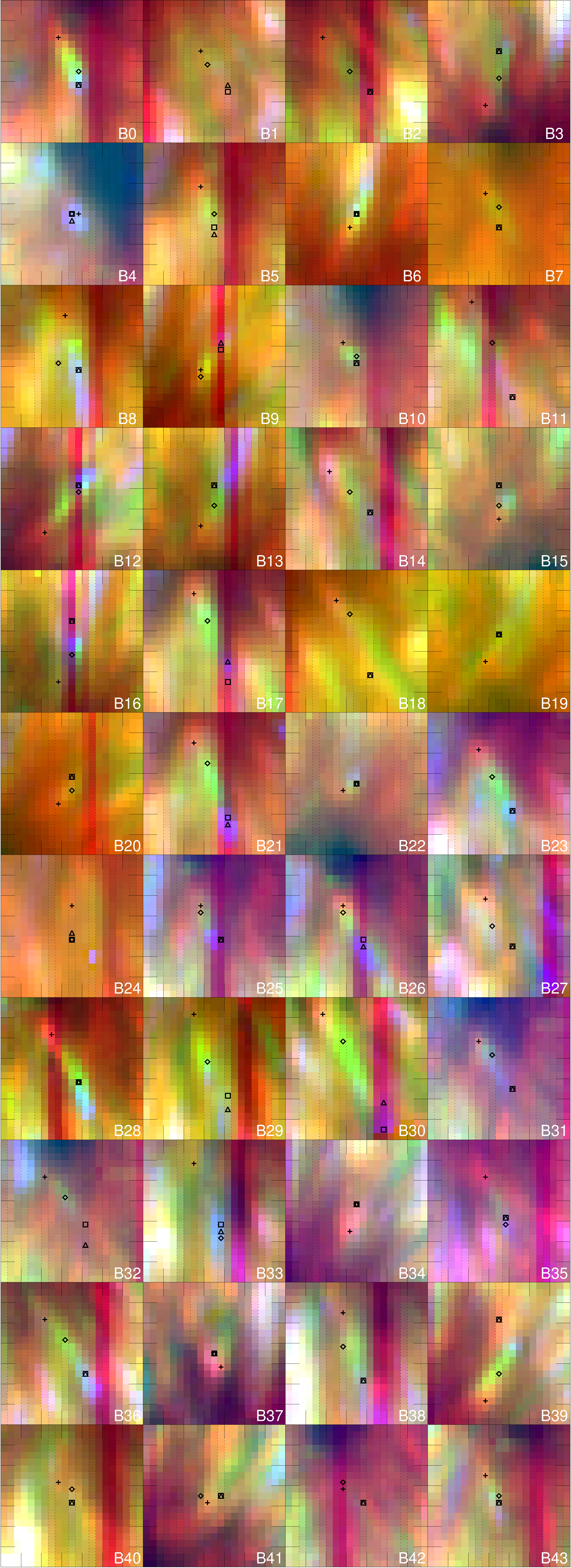} 
\caption{\label{fig:rainbows_30} All PMJs from dataset B. The layout is identical to that of Fig. \ref{fig:rainbows_29}, see its caption for details.}
\end{figure}
%=======================================

PMJs in our observations adhere to the same type of morphology as when previously studied in a variety of works, and they present as elongated brightenings that appear and disappear on time scales of minutes or less than a minute. This overall appearance also holds true when PMJs are viewed in multiple wavelength channels simultaneously. 
As mentioned, PMJs often display a rainbow-like signature in composite-channel RGB images, which was used as another aid both in discovering and identifying PMJs, but it was not used as a strict criterion in order to make a positive identification. 
Figures \ref{fig:rainbows_29} and \ref{fig:rainbows_30} display all 
of the studied PMJs in composite RGB images for datasets A and B, respectively. 
Many PMJs exhibit a rainbow signature with a connected red-to-green-to-blue elongated structure that reaffirms the observation that PMJs typically exhibit a signal in channels associated with temperatures that correspond to the chromosphere and up to the TR, suggesting that PMJs range through such temperatures. 

PMJs do not always exhibit a clear rainbow pattern. In some instances, this is due to the PMJ in question, which simply does not exhibit a signal in the relevant channel. Positing that a PMJ originates in the chromosphere (the validity of which is discussed in Sect. \ref{sec:discussion}), less energetic ones may not reach the relevant temperature (or height) to which the channel corresponds. In other cases, it is due to a failure in capturing the relevant signal. Another source for absent rainbow patterns may result from the automated mixing in the RGB image resulting from the three diagnostic images, which may not always yield favourable visual results. Lastly, in several images, the shadow of the IRIS spectrograph slit is also evident, usually as a purplish line in the images, and it may obscure possible rainbow patterns. 
Despite a number of PMJs that do not exhibit a rainbow pattern, a majority of the detected PMJs can reasonably be said to display such a pattern for both our dates of observations.  

\subsubsection{PMJ sizes}\label{sec:pmj_sizes}

We measured a lower bound on PMJs lengths that are based on two sets of sampling positions. The first is the sampling position of \efft\ and the second is the sampling position of the \siiv\ lines. The inferred lengths are the straight-line distances between these two sampling positions for any given PMJ. These sampling positions are located at the nominal peak-intensity pixels at the nominal peak-intensity time. As such, the sampling positions do not measure the maximum extents of our PMJs, as there is often still some appreciable signal away from the given peak-intensity pixel. However, \siiv\ line core intensities typically dropped dramatically with increased pixel distance from the peak-intensity pixel. Ultimately, the presented lengths must therefore be considered a lower bound on projected PMJ lengths when measured from the chromosphere to the TR. See both the overview in Figures \ref{fig:all_pmjs29} and \ref{fig:all_pmjs30} and the PMJ RGB-images in Figs. \ref{fig:rainbows_29} and \ref{fig:rainbows_30} for reference.

The mean lower-bound lengths with associated standard deviations for our studied PMJs were found to be $(2204 \pm 195.0)$ km and $(1940 \pm 138.2)$ km for datasets A and B, respectively.

Multi-channel PMJ width estimates necessitate measurements in IRIS SJI 2796 images since PMJs usually display their midsection in this channel. However, widths of most PMJs fell close to the pixel resolution of the \mgii\ SJI images, typically of the order of 1-3 pixels, meaning potential width distributions would be of dubious value at best. We thus constrain ourselves to the observation that our PMJs have typical widths of 243 km - 729 km, corresponding to the widths of 1 - 3 pixels in the \mgii\ SJI images.

\subsubsection{Overview of PMJ responses in different spectral lines}\label{sec:spectral_overview}

%======================================= fig 6
\begin{figure*}[!ht]
\centering
\includegraphics[width=17.5cm]{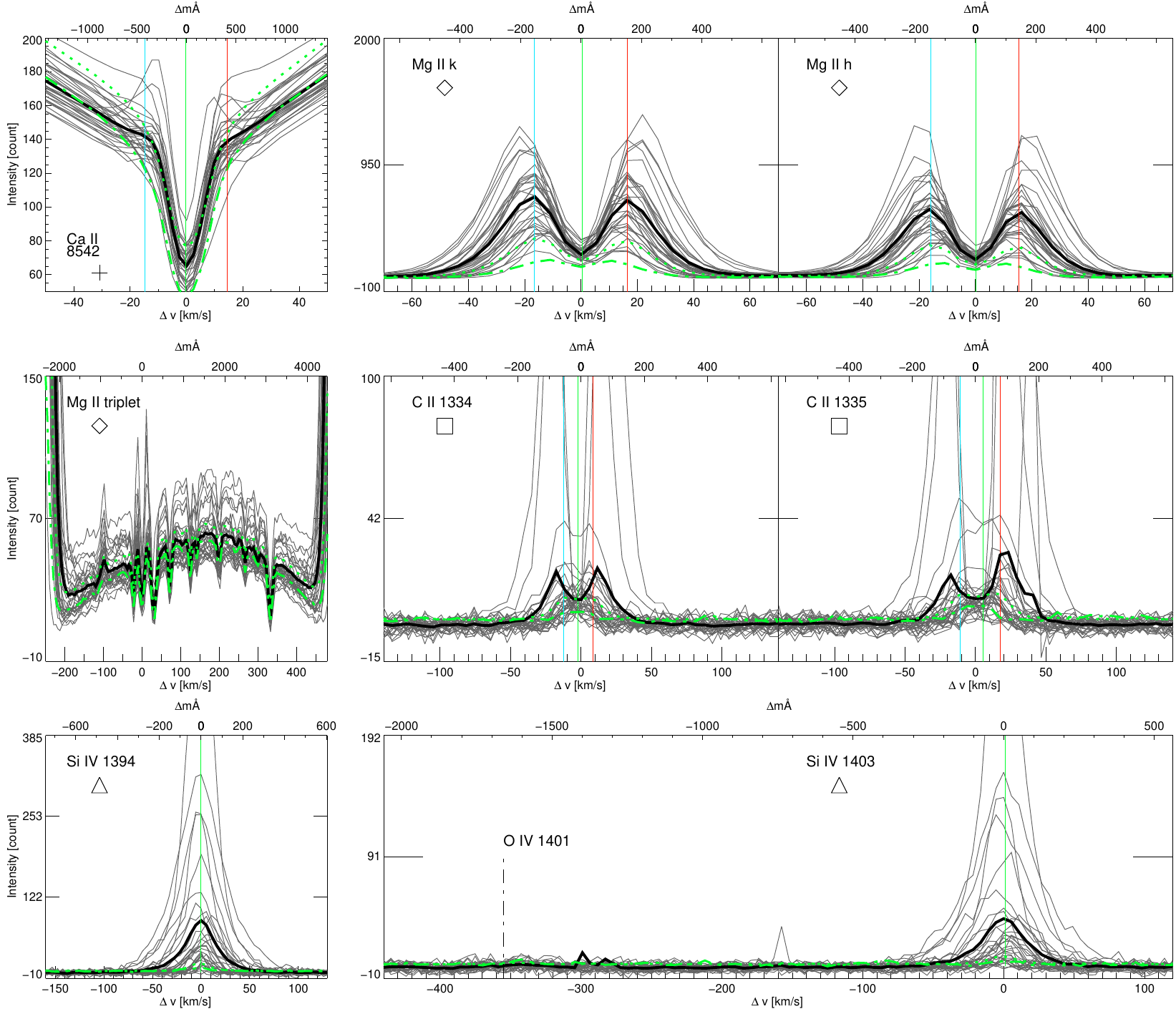} 
\caption{\label{fig:29th_all_profiles} Spectral signatures of PMJs and reference average profiles in various spectral ranges from dataset A. The number of PMJ events is N = 33. Shown in panels left to right and top to bottom are profiles of the \efft\ line, the \mgii\ k and h lines, the \mgii\ triplet, the \cii\ 1334 \AA\ and 1335 \AA\ lines, and the \siiv\ 1394 \AA\ and 1403 \AA\ lines. Each 
panel is also labelled with the relevant profile name and the symbol that each spectral line is associated with when plotting the position from which the profile is sampled in other figures (see for instance Figures \ref{fig:rainbows_29} and \ref{fig:rainbows_30}). Furthermore, the spectral position of the \oiv\ line in the panel of the \siiv\ 1403 \AA\ line is also indicated by a vertical dash-dotted line. The different line styles in the panels, which are associated with the different spectral lines mentioned, indicate the different individual and average line profile and denote the following: grey solid lines, the line profiles of all individual PMJs detected in the observations; black solid line, the average profile of all detected PMJs; green-dotted line, the average line profile across all pixels in the observations; and green-dash-dotted line, the average line profile in the penumbra of AR12533. For all studied PMJ profiles, the mean positions of line core positions (vertical solid green lines) and those of red peaks (vertical solid red lines) or blue peaks (vertical solid blue lines) for those spectral lines where relevant are
also marked in each panel. 
}
\end{figure*}
%=======================================

%======================================= fig 7
\begin{figure*}[!ht]
\centering
\includegraphics[width=17.5cm]{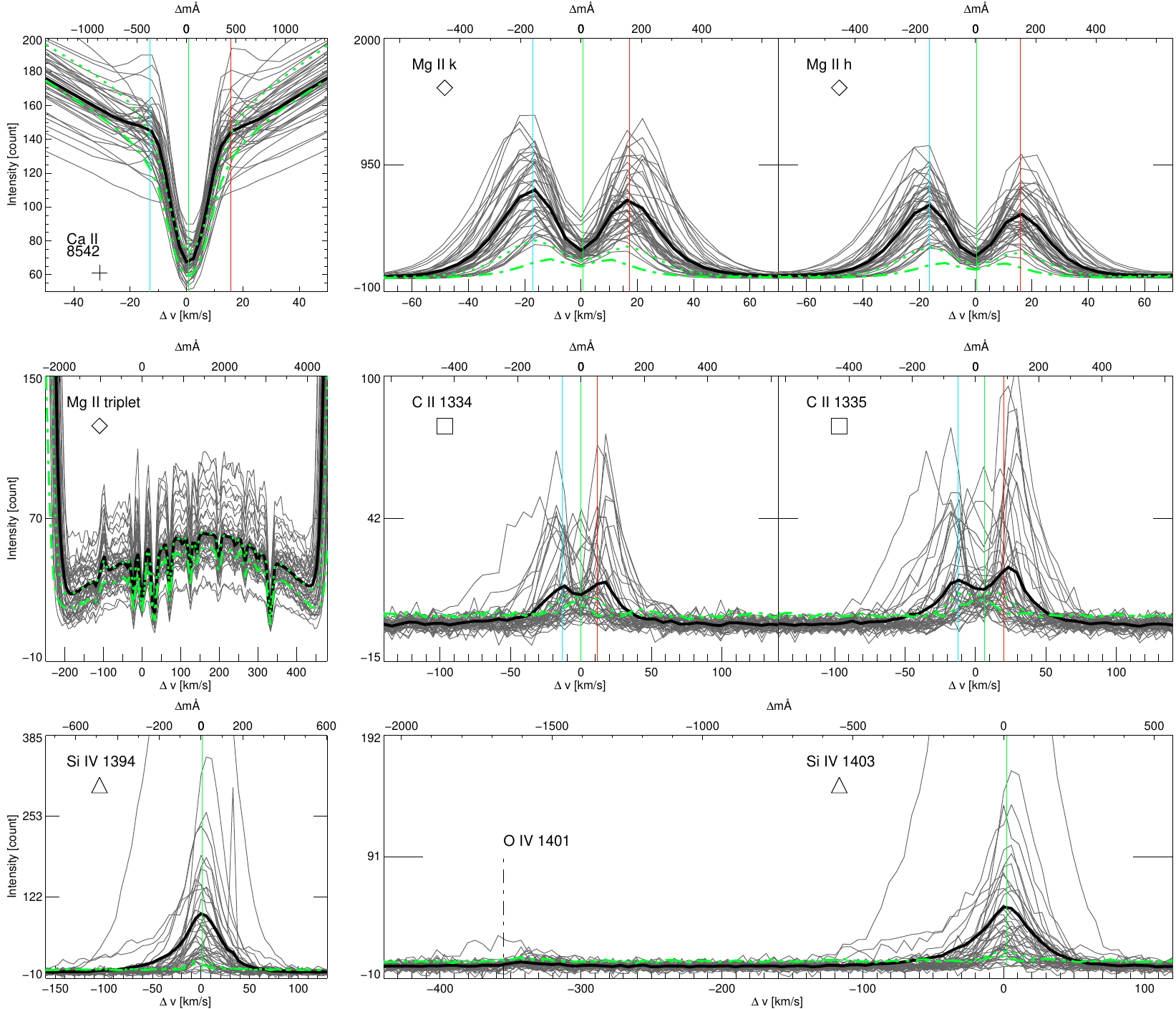} 
\caption{\label{fig:30th_all_profiles} Spectral signatures of PMJs and reference average profiles in various spectral ranges from dataset B. The number of PMJ events is N = 44. The layout is identical to that of Figure \ref{fig:29th_all_profiles}, see its caption for details.}
\end{figure*}
%=======================================

\begin{table}[!t]
  \begin{center}
\begin{threeparttable}
  \caption{Number of occurrences of distinct PMJ signal in different spectral lines.}
  \begin{tabular}{l l l l} 
Dataset A, $\text{N}_\text{total}$ = 33  \\
\hline 
Spectral line(s) & N & \%\ \\
\hline
  \efft\                        & 29 &  (88\%)\\
  \mgii\ h \& k         & 32 &  (97\%)\\
   Wings of \mgii\ $2798.75$ \AA\ \& $2798.82$ \AA\ \\
  Triplet Blend & 7       &     (21\%) \\  
  \cii\          1334 \AA\ \& 1335 \AA\ & 23 &  (70\%)\\
  \siiv\ 1394 \AA\ \& 1403 \AA\         & 25 &  (76\%)\\
\hline \\
Dataset B, $\text{N}_\text{total}$ = 44 \\
\hline
Spectral line(s) & N & \%\ \\
\hline
  \efft\                        &  40 & (91\%) \\
  \mgii\ h \& k &  42 & (95\%) \\
   Wings of \mgii\ $2798.75$ \AA\ \& $2798.82$ \AA\ \\
  Triplet Blend & 5  &  (11\%) \\
  \cii\ 1334 \AA\ \& 1335 \AA\          & 32 &  (73\%) \\
  \siiv\ 1394 \AA\ \& 1403 \AA\         &  37 & (84\%) \\
\hline
\end{tabular}
\label{tab:PMJ_profile_stats}
\end{threeparttable}
\end{center}
\end{table}

Figures \ref{fig:29th_all_profiles} and \ref{fig:30th_all_profiles} show spectral profiles for all of our studied PMJs. The average event profiles and average profiles for the penumbra and the entire FOV for each date observation for all given lines are also shown. We note that \halpha\ line profiles for PMJ dark features are presented separately in Sect. \ref{sec:results_halpha}.  

PMJs may exhibit signals in all of the lines. However, the studied PMJs display enhancement more readily in some of the lines than in others.

The figures highlight that there is a large spread in the profiles for all channels, but for both datasets it is evident that PMJs often display strong signals in the \efft\ and the \mgii\ h and k lines compared to the penumbral average, while enhancement in other lines is somewhat less common, especially the \mgii\ triplet lines. We note that \cii\ 1334 \AA\ \& 1335 \AA\ typically exhibit double peaks when in emission, but also some single peaks. When in emission, the \siiv\ 1394 \AA\ \& 1403 \AA\ lines always present as single peaked and near-Gaussian. 

There exists one example of a PMJ exhibiting enhancement in the \oiv\ line in dataset B (see Fig. \ref{fig:30th_all_profiles}). 
This event also 
corresponds to the strongest event in terms of intensity for the \siiv\  lines, and with very defined profiles with strong intensities for the other lines (but with no emission in the \mgii\ triplet). 
This may indicate that the specific PMJ event may be heated to temperatures far beyond earlier observed examples. This specific PMJ is presented in Sect. \ref{sec:results_oiv}, albeit briefly, as a single event should not be overanalysed nor overemphasised.

\begin{table*}[!t]
  \begin{center}
\begin{threeparttable}
  \caption{\efft\ line diagnostic values for our studied PMJs.}
  \begin{tabular}{l l l l} 
Dataset &         line core offset & blue peak offset & red peak offset   \\
 & [km/s]  & [km/s]  & [km/s]  \\
\\
A, N = 33 & $  -0.17\pm 0.00$ {\small (N=33)} &$-14.80\pm 0.02$ {\small (N=16)} &      $ 14.55\pm 0.03$ {\small (N=12)} \\
\hline
\\
B, N = 44 & $  0.81\pm 0.00$ {\small (N=44)} &$-12.93\pm 0.01$ {\small (N=20)} &      $ 15.78\pm 0.02$ {\small (N=15)} \\
\hline
\end{tabular}
\label{tab:8542_profile_stats}
\end{threeparttable}
\end{center}
\end{table*}

Table \ref{tab:PMJ_profile_stats} summarises the number of PMJs that exhibit clear signals in the various channels studied. A clear signal means that a given line profile shows an enhancement when compared to nearby pixels in space and time, and it is thus based not only on a comparison to the penumbral average for a given line since profile intensities can vary substantially with regards to position throughout the sunspot. This investigation was 
performed visually to give an estimate of the PMJ signal frequency in the different channels. 

\subsection{PMJs in the \efft\ line}\label{sec:results_8542}

PMJs have been described in the \efft\ line in the literature by  
\cite{2013ApJ...779..143R},
\cite{2017A&A...602A..80D}, 
\cite{2015ApJ...811L..33V}, 
\cite{2019ApJ...870...88E}, 
and 
\cite{2019ApJ...876...47B}. 
We investigated the line-core position and the positions of any present peaks in the \efft\ line for all of our investigated PMJs. In order to do this, we employed a simple spline interpolation of the line profiles, followed by an approach in which subsections of the profile were iteratively inspected for local peaks and minimums. This process is akin to the one previously employed in 
\cite{2017A&A...602A..80D}. 
This approach always identifies line cores and those inner line wing enhancements that have clear peaks. Inner line wing enhancements that are merely somewhat enhanced compared to the average penumbral \efft\ line profile and that may be discernible by eye were not selected. A consistent way to select offset positions for these is non-trivial, and we chose to focus on the cases where the enhancements correspond to clear peaks to avoid ambiguity. 

The values for line core positions as well as the blue and red peak positions are tabulated in Table \ref{tab:8542_profile_stats}. The average values are given 
together with the respective sample sizes. It is important to note that the number of line core and peak positions does not necessarily match the number of PMJs for each category; especially identifiable peaks are not present for all PMJs. Also, we note that the given uncertainties are standard errors, which are based solely on the sample sizes and the standard deviations of the PMJ subsets, rather than instrumental or noise-based uncertainties, as these are expected to be subsumed by the former.

For both sets of observations, the appearance of our PMJ \efft\ line profiles is, in general, in agreement with those in the literature. The average profiles for both dates are enhanced in both the line core as well as the inner blue and red line core compared to the penumbral average. See Figs. \ref{fig:29th_all_profiles} and \ref{fig:30th_all_profiles} again. The average \efft\ PMJ line profiles exhibit the distinct blue-over-red asymmetry in the inner line core, which is seen the most distinctly for dataset A, and in general there are more distinct blue peaks that are distinguishable in individual profiles. The latter can be seen from the sample sizes for the automatically detected blue and red peak positions cited in Table \ref{tab:PMJ_profile_stats}. The overabundance of blue peaks over red peaks is especially in good agreement with the statistical study 
performed in 
\cite{2017A&A...602A..80D}.

The specific values (see Table \ref{tab:8542_profile_stats}) for the line core and the blue and red peak offsets for the PMJ line profiles are also in general agreement with the literature. The line core Doppler offsets are close to zero, and there is less than $1$ \kms\ of apparent shift from the nominal line core for both days of observation. The blue and red peak offsets are found at near-equal positions on the blue and red side of the line core for both dates of observations. The values found for these positions are also in general agreement with the values found for the large dataset of automatically detected PMJs in 
\cite{2017A&A...602A..80D}. 
Here the average line core, blue- and red-peak offsets in the \efft\ line were found to 
be $0.16$ \kms, $-12.12$ \kms\, and $11.95$ \kms\, respectively\footnote{These values were corrected for a conversion error in \cite{2017A&A...602A..80D} 
that stemmed from a wrong constant in a programming routine that converted values from units of Ångstrom to \kms. As given here, they also follow the sign convention in this paper that negative values are on the blue side of the line-core. 
}.

\subsection{PMJs in the \mgii\ h \&\ k line pair}\label{sec:results_mgii}

\begin{table*}[!ht]
  \begin{center}
\begin{threeparttable}
  \caption{Mean values and standard errors for different line diagnostics of the \mgii\ h \& k lines for detected PMJs from datasets A and B. }
\begin{tabular}{l l l l} 
\hline
 Dataset/&Core offset$^\text{a}$        & Peak$^\text{a}$               & Avg. Doppler$^\text{a}$ \\
line  &                         & separation    & shift          \\
 & $\Delta \text{v}_\text{x3}$ & $\left( \lvert \Delta \text{v}_\text{x2v} \rvert+ \lvert \Delta \text{v}_\text{x2r} \rvert \right)$ & $\frac{1}{2}\left(\Delta \text{v}_\text{x2v} + \Delta \text{v}_\text{x2r} \right)$  \\
 & [\kms]  & [\kms] & [\kms] \\
\hline
\\
A (N = 33)\\
\hline
\mgii\ k & $  0.47\pm 0.25$   &$ 33.14\pm 0.35$ &$ -0.02\pm 0.18$  \\
\mgii\ h & $  0.19\pm 0.25$  &$ 31.25\pm 0.36$ &$ -0.23\pm 0.18$  \\
\hline
\\
B (N = 44) \\
\hline
\mgii\ k & $  0.65\pm 0.32$   &$ 34.16\pm 0.51$ &$ 0.08\pm 0.26$ \\
\mgii\ h & $  0.36\pm 0.32$  &$ 32.39\pm 0.44$ &$ -0.27\pm 0.22$  \\
\hline
\end{tabular}
\begin{tablenotes}
\item[a] Where `x' in the formulae may be k or h, as relevant for the given row.
\end{tablenotes}
\label{tab:PMJ_mgii_stats}
\end{threeparttable}
\end{center}
\end{table*}

We have presented the average spectral profiles in the \mgii\ line pair for our two sets of PMJ observations in Figs. \ref{fig:29th_all_profiles} and \ref{fig:30th_all_profiles}. In general, PMJs do not exhibit distinct features other than the double peak structure, which is typical for both lines in penumbrae, though they do exhibit a marked increase in average peak-intensities compared to the average \mgii\ line pair profile of the sunspot penumbra in the observations of both dates.
Values for the various diagnostics are given in Table \ref{tab:PMJ_mgii_stats}.

\subsubsection{\mgii\ h \& k line core shifts}\label{sec:results_mgii_line_core}

The line core shifts of the \mgii\ h \& k lines, $\Delta \text{v}_\text{k3}$  and $\Delta \text{v}_\text{h3}$, both strongly correlate with the line of sight velocity at the formation height of the $\text{h}_\text{3}$ and $\text{k}_\text{3}$ features at optical depth unity. The correlation coeffiecient was shown to be close to unity in \cite{2013ApJ...772...90L}, 
and these shifts therefore hold great immediate diagnostic value and constitute the 
most reliable of the \mgii\ diagnostics, probing either PMJs or the atmospheric conditions in their immediate surroundings. The height of optical depth unity for both $\text{h}_\text{3}$ and $\text{k}_\text{3}$ is close to that of the very upper chromosphere, and the Doppler-velocity offsets of the $\text{h}_\text{3}$  and $\text{k}_\text{3}$ features can therefore be taken as diagnostic proxies for the line of sight velocities at this height.
Table \ref{tab:PMJ_mgii_stats} gives the Doppler-velocity offsets of the $\text{h}_\text{3}$ and $\text{k}_\text{3}$ features for our sample positions in the \mgii\ h \& k lines for our detected PMJs for both our sets of observations. The measured values are consistent across all four given values, and they are all close to zero with an absolute value range $\le 0.97$ \kms. 

\subsubsection{\mgii\ h \& k line peak separations}\label{sec:results_mgii_peak_seps}

We define the peak separations of the \mgii\ h \& k line as the absolute sum of the Doppler shifts of the two peaks in each \mgii\ line, $\lvert \Delta \text{v}_\text{x2v} \rvert+ \lvert \Delta \text{v}_\text{x2r} \rvert $ (where `x' may be k or h). 
\cite{2013ApJ...772...90L} and \cite{2013ApJ...778..143P} show that this quantity correlates with the difference between the maximum and minimum of the atmospheric line of sight velocity between the mid-chromospheric formation heights of the peaks and heights above. The relation is less reliable than the previous diagnostics, with an average correlation coefficient of 0.57, which was inferred for the two lines. Further, it was shown that a temperature maximum in the lower chromosphere can also cause wider peak separations, which means that a discussion regarding this diagnostic must be more nuanced, especially given that PMJs are energetic events that are hypothesised to form in the chromosphere. 
Table \ref{tab:PMJ_mgii_stats} gives the peak separations for the h \& k line of our \mgii\ sample positions of our detected PMJs for both dates of observations.

All values for the peak separations for both dates are rather consistent, with values slightly higher for the k line than for the h line for both dates of observations, and the separations are overall higher for the observations of dataset B. Overall, the peak separation values are of the order of $\approx 30$ \kms. This may indicate a maximum velocity difference between the mid-chromosphere and above in this range, assuming there is no temperature maximum in the lower chromosphere. This assumption, however, is challenged in Sect. \ref{sec:discussion} due to other diagnostics indicating that there may indeed exist such a temperature maximum.

\subsubsection{\mgii\ $\text{h}_2$ \& $\text{k}_2$ peak shifts}\label{sec:results_mgii_avg_peak_doppler}

\cite{2013ApJ...772...90L} and \cite{2013ApJ...778..143P} show that the average Doppler shifts of the \mgii\ $\text{h}_2$ and $\text{k}_2$ peaks correlate with the line of sight velocity at the optical depth $\tau = 1$ height for the peaks in each line. Further, the $\text{h}_2$ and $\text{k}_2$ peaks are shown to form in the mid-chromosphere, about 1 Mm 
below their line cores ($\text{h}_\text{3}$ and $\text{k}_\text{3}$). The average Doppler shifts of the peaks thus correlate to the line of sight velocity at this height. The average of the correlation coefficients between average peak Doppler shifts for the two lines and the line of sight velocity at their formation heights is $0.66$, and thus this diagnostic is somewhat less reliable than the one for the line core Doppler shifts, but it still provides a strong correlation.

From Table \ref{tab:PMJ_mgii_stats} we see that the averaged peak Doppler shifts are all close to zero \kms\ in magnitude; the majority of which are negative, indicating an upflow. With standard errors included, we find the very small upper absolute value of $\le 0.49$ \kms\ for the average peak Doppler shifts across all PMJs. Overall, negative values may indicate a possible subtle upflow in the lower chromosphere, though in which case it is close to zero in value.

\label{sec:results_mgii_int_as}
%======================================= fig 8
\begin{figure*}[!h]
\centering
\includegraphics[width=18cm]{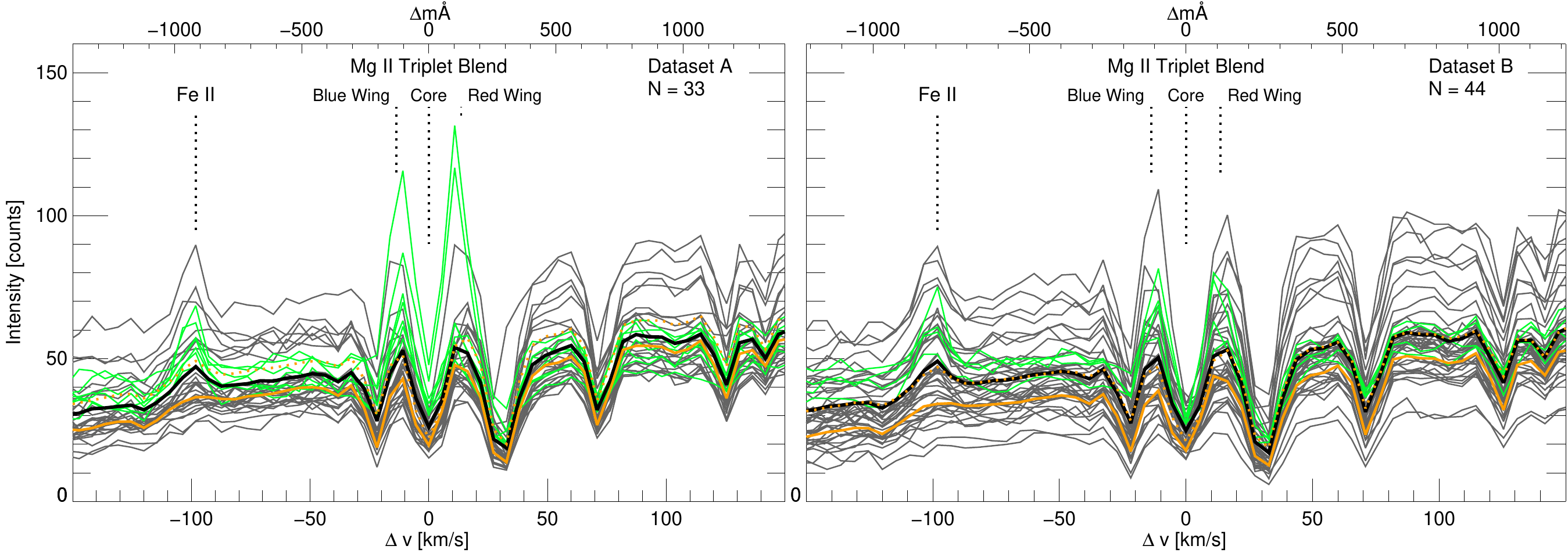} 
\caption{PMJ profiles in the wavelength region between the \mgii\ h \& k lines for datasets A (left) and B (right). All PMJ profiles that were sampled at sampling positions given in Figure \ref{fig:all_pmjs29} and \ref{fig:all_pmjs30} are shown (marked with green diamond symbols), coinciding with the \mgii\ h \& k sampling positions. We show all individual profiles (grey solid line) together with those profiles that were identified to show emission in the triplet blend wings (green solid line). The mean profiles for all PMJ profiles (black solid line), the penumbra (yellow solid line), and the full field of view (yellow-dotted line) are also shown. \label{fig:pmj_triplets} }
\end{figure*}
%=======================================

\subsection{PMJs in the \mgii\ triplet region}\label{sec:results_triplet}

The \mgii\ triplet spectral lines are located at wavelength positions $2798.75$ \AA, $2798.82$ \AA,\ and $2791.60$ \AA\ and provide diagnostics on the lower chromosphere and the photosphere. Unfortunately the $2791.60$ \AA\ position was not covered by our IRIS observations, and its response could therefore not be studied. The two triplet lines at $2798.75$ \AA\ and $2798.82$ \AA\ are not distinguishable as separate lines with the resolution of IRIS, and therefore they appear as a blend. We refer to this blended line as the triplet blend going forward. We studied the response of the \mgii\ triplet blend for our PMJ events and present our findings below.

\cite{2015ApJ...806...14P}
found that emission in the \mgii\ triplet blend line core is associated with a steep temperature increase in the lower chromosphere. It was found that the \mgii\ triplet blend typically appeared in emission when there was an increase in temperature of more than 1500 K in the lower chromosphere. In the case that heating occurs deeper down in the atmosphere, at photospheric heights, it was found that typically only the wings of the \mgii\ triplet blend are enhanced whilst the core itself does not exhibit emission. 
To our knowledge, PMJs have not been previously shown to exhibit intensity enhancement in any of the \mgii\ triplet lines. We do not find any PMJs for which we see emission in the core of the triplet blend line, but we do find multiple occurrences of emission in its wings, indicating heating at photospheric heights. 

The number of PMJs that show significant emission in the \mgii\  triplet blend wings as determined by eye are tallied in Table \ref{tab:PMJ_profile_stats}, as presented previously. For reference purposes, Fig. \ref{fig:pmj_triplets} shows an overview of PMJ profiles in the wavelength region between the \mgii\ h \& k lines. Clear emission in the triplet blend wings is evident for several PMJs in the different wavelength positions. These events are indicated and correspond to those tallied in Table \ref{tab:PMJ_profile_stats}.
It must be noted that profiles with enhanced wings in the triplet blend were selected by comparing the wings with the overall intensity of the wavelength region for each given PMJ. As can be seen in Fig. \ref{fig:pmj_triplets}, the profiles of PMJs cover a rather large intensity range. Absolute intensities in PMJs selected as showing an enhancement in the triplet blend wings may in some cases have lower intensity in these positions than other PMJs that were not selected. However, the overall intensity of the spectral range between the \mgii\ k and h lines can vary quite substantially, which may often be attributable to far-wing emission from these two lines. As such, only profiles that show clear enhancement in the triplet blend wings as compared to the surrounding emission were selected, and they were not selected solely on a criteria of absolute intensity levels. We therefore also encourage the reader to specifically inspect the triplet blend plots for individual PMJs in the peak brightness PMJ videos for both dataset A 
(\href{https://www.mn.uio.no/astro/personer/vit/rouppe/movies/drews_2020_appa_video1.mp4}{Video 1}) and B 
(\href{https://www.mn.uio.no/astro/personer/vit/rouppe/movies/drews_2020_appa_video2.mp4}{Video 2}), which are included in the online material. A sample frame, Fig. \ref{fig:vid_example} with an accompanying caption, is given in Appendix \ref{sec:appendix_videos}. The individual triplet blend plots in the videos specifically highlight the marked appearance of emission in the triplet blend wings as compared to its far wings.

We also note that we observe emission in a line adjacent to the two observed \mgii\ triplet lines that is situated at $2797.868$ \AA. 
This line lies between the \mgii\ triplet blend and the \mgii\ k line and 
was recently identified as a \feii\ line in
\cite{2015ApJ...811..139T}. 
The identification of this line is uncertain, however 
\citepads[see ]{2019ApJ...878..135K}, 
and we limit ourselves to reporting the observation that this line is typically in emission whenever the wings of the triplet blend are also in emission. 

We computed the mean values of the intensity at the two \mgii\ triplet blend wing positions for each PMJ and normalised these to the corresponding penumbral mean. The triplet blend wings are typically mildly enhanced compared to the penumbral mean, with normalised mean values close to 1.2 for the two datasets,  with rather Gaussian-like distributions. We see several outliers towards higher relative intensities corresponding to events in which the triplet blend wings are in clear emission.

\subsection{PMJs in the \cii\ line pair}\label{sec:results_cii}

\begin{table*}[!ht]
  \begin{center}
\begin{threeparttable}
  \caption{Average values and their standard errors for different line diagnostics of the \cii\ line pair for different groups of PMJs for datasets A and B are given.}
\begin{tabular}{l l l l l l} 
\hline
Dataset/ &  Line core  & Line  & Line Int.  & Single & Double  \\
line & offset & width    & Ratio & Peaks  & Peaks \\
& & $\text{W}_\text{FWHM}$ & $\text{R}_\text{I} = \frac{\text{I}_\text{1335}}{\text{I}_\text{1334}}$  &   & \\
&   [\kms] & [\kms] & [ratio] & [nr] &  [nr] \\
\hline
\\
A  (N = 33) \\
\hline
\\
\cii\ 1334 \AA\ & $  -2.14\pm 0.67$ {\small (N=22)} &$ 31.33\pm 12.63$ {\small (N=22)} &       & 13 & 9 \\
\cii\ 1335 \AA\ & $  5.20\pm 0.67$ {\small (N=18)} &$ 41.71\pm 11.30$ {\small (N=18)} & & 7 & 12 \\
\cii\ line pair& & &       $ 1.20\pm 0.08$ {\small (N=18)}  \\
\hline
\\
B (N = 44)  \\
\hline
\\
\cii\ 1334 \AA\ & $  -0.32\pm 0.95$ {\small (N=29)} &$ 43.19\pm 14.46$ {\small (N=29)} & & 8  & 21  \\
\cii\ 1335 \AA\ & $  6.13\pm 0.81$ {\small (N=31)} &$ 51.36\pm 16.06$ {\small (N=31)} & & 6 & 25 \\
\cii\ line pair& & &       $ 1.19\pm 0.07$ {\small (N=28)}  \\
\hline
\end{tabular}
\label{tab:PMJ_cii_stats}
\end{threeparttable}
\end{center}
\end{table*}

We have presented \cii\ line pair profiles for all of our detected PMJs in Figs. \ref{fig:29th_all_profiles} and \ref{fig:30th_all_profiles}. In those cases in which there is an actual enhancement in the line pairs, the profiles present with a double-peak profile in the majority of cases. There exist a few select cases for which we see single-peak profiles.
We base our analysis of the \cii\ line pair on 
\cite{2015ApJ...811...80R}, 
\cite{2015ApJ...811...81R}, and 
\cite{2015ApJ...814...70R} 

The \cii\ line pair can, in principle, exhibit quite different types of line profiles, with both single, double, and more peaks, and therefore a proper definition of the line core must be established that covers these different cases. We define the line core as in the previously cited papers; for single peaks, we define it as the local maximum that is close to the nominal line centre and for double peaks as the local minimum of the intensity reversal between the two peaks. We do not observe more than two peaks in our investigated profiles.

For the different diagnostics, we utilise the offset values for the line core and the positions of blue or red peaks, when these are present. Our automated method for finding local maxima and minima usually determines the positions of these profile features to a good approximation when validated by eye, although misidentifications are far more common than for the \mgii\ h \& k lines. This is due to the \cii\ line pair observations being far noisier than in the \mgii\ lines, and because the \mgii\ h \& k lines always present with a well-defined double peaked profile for our sampled PMJs. In contrast, the \cii\ line pair profiles only exhibit appreciable enhancement in a subset of cases, if at all, and in these cases, they also suffer from the poorer signal-to-noise ratio. As such, the diagnostic values inferred from our investigation of the \cii\ line pair should, in general, be viewed as less reliable than those for the \mgii\ h \& k lines. However, given the appreciable number of PMJs that exhibit clear enhancement, our mean values for the various diagnostics should still be viewed as valid for a qualitative interpretation. Further, the fact that we have observed on two different dates and find similar values for our diagnostics on the two dates strengthens their validity. 

It is worth mentioning that the \cii\ 1335 \AA\ line is comprised of a blend of two lines, namely the stronger $1335.708$ \AA\ and the weaker $1335.663$ \AA\ component. In general, this means that the diagnostics presented for the \cii\ line pair are less reliable for the \cii\ 1335 \AA\ line than for the 1334 \AA\ line. 

The diagnostics described below were inferred from those PMJ profiles that have a maximum intensity in the $\pm 50$ \kms\ offset region of the nominal line core of the given line that is equal to or greater than that of the mean profile in the same line over the entire field of view for the given dataset. This was done to limit the samples to those profiles which show an enhancement that is actually appreciable as well as to not include values that are either very distorted by noise or that are altogether spurious. 
All inferred mean diagnostics values for the investigation of the \cii\ line pair for datasets A and B are given in Table \ref{tab:PMJ_cii_stats}.

\subsubsection{\cii\ line-pair line core shifts}\label{sec:results_cii_line_core_shifts}

The line cores of the \cii\ 1334 \AA\ and 1335 \AA\ are formed approximately just below the TR \citep{2015ApJ...811...81R}, and the line core offsets of the two lines thus offer a diagnostic to approximate the line of sight velocity at this height. In 
\cite{2015ApJ...811...81R}, 
it is shown that for the \cii\ 1334 \AA\ and 1335 \AA\ lines, there is a good correlation between the offsets in their line cores to the vertical velocity at line core formation heights, with correlation coefficients of 0.69 and 0.63, respectively. 

From Table \ref{tab:PMJ_cii_stats}, we see that values for the \cii\ 1334 \AA\ and 1335 \AA\ line core shifts are consistent across the dates of observations, both having the same sign for each line and being on the same order of magnitude. The 1334 \AA\ line has a negative line core offset for both dates, with a maximum absolute value of $\le 2.81$ \kms\ when considering standard errors. The 1335 \AA\ line has a positive sign for both mean values for the two dates of observations, with a maximum absolute value of $\le 6.94$ \kms\ when considering standard errors.

The line core offsets with conflicting signs for the two lines indicate a low overall line of sight velocity close to the TR that is less than 7 \kms\ in absolute value. Even though both \cii\ lines form close to the TR, the stronger 1335 \AA\ forms slightly higher than the 1334 \AA\ line. Thus, in principle, the difference in sign for the core offsets may indicate a small difference in the line of sight velocity at these slightly different heights.

\subsubsection{\cii\ integrated intensity line-pair ratio and number of peaks}\label{sec:results_cii_integrated_intensity_ratio}

The integrated intensity ratio of the \cii\ 1334 \AA\ and 1335 \AA\ lines can be used as a measure to determine whether it is formed under optically thick or possibly under optically thin conditions. The line intensity ratio is also correlated to the number of peaks present in the given profile, and the number of peaks gives insight into the type of source function that produces their line profiles. 
We follow the definition of the integrated line ratio of the \cii\ 1334 \AA\ and 1334 \AA\ lines in  
\cite{2015ApJ...811...80R}, 
such that we define the intensity ratio as $\text{R}_\text{I} = \frac{\text{I}_\text{1335}}{\text{I}_\text{1334}}$. Here, the intensities $\text{I}_\text{1334}$ and $\text{I}_\text{1335}$ are the intensities of the two lines integrated over the region of $\pm 20$ \kms\ around the nominal line core for each given line.  
It was shown that this ratio can, in principle, take any value in the case of optically thick line formation, but it has the value $\text{R}_\text{I} = 1.8$ for the case of optically thin line formation. This means that $\text{R}_\text{I} = 1.8$ is compatible with optically thin line formation, but this does not prove it, whilst any other value indicates optically thick line formation. Furthermore, it was shown in modelling that the value of $\text{R}_\text{I}$ is typically lower for double peak profiles, with values around $1.4$, whilst single peak profiles have typical values around $1.7$ 
\citep{2015ApJ...811...81R}. 

Table \ref{tab:PMJ_cii_stats} gives the values of the integrated line-pair ratio for the two dates of observations in column five together with its associated standard error. The values are consistent and overlap, given their standard error and give a range of ($1.12$ -- $1.28$). The values thus indicate optically thick line formation. The values are also closer to the mean value of $\text{R}_\text{I} = 1.4$ expected for double peak profiles than  the one for single peak profiles. This is congruent with the fact that we find a majority of double peak profiles when there is enhancement in either of the lines. The number of single and double peak profiles are given in Table \ref{tab:PMJ_cii_stats}. For all but the case of the \cii\ 1334 \AA\ line for dataset A, we find a greater number of double peaks than single peaks when there is any clear enhancement in the given line. 

Single peak profiles in both the \cii\ 1334 \AA\ and 1335 \AA\ lines are formed when the source function of the given line increases monotonically up to the height where the line core has optical depth unity. Double peak profiles are formed when there is a local maximum in the source function deeper down in the atmosphere, and if the source function above this local maximum decreases in value with increasing height until optical depth unity for the line core is reached. The \cii\ 1335 \AA\ line overall tends to have a higher number of double peaks than its counterpart, since its line core forms overall a little higher than that of the \cii\ 1334 \AA\ line, and thus there is a greater likelihood that the 1335 \AA\ line core forms at a greater height than a potential local maximum in its source function compared to the 1334 \AA\ line \citep{2015ApJ...811...80R, 2015ApJ...811...81R,2015ApJ...814...70R}.  
We see the latter effect exemplified in Table \ref{tab:PMJ_cii_stats}, as there is a greater prevalence of double peaks in the 1335 \AA\ line than for the 1334 \AA\ line for both dates of observation. Further, as we have a greater number of double peaks than single peaks overall, we can reasonably conclude for a majority of cases in which there is an enhancement in the \cii\ line pair that there is a local maximum in the source function below the formation height of the line cores of the two lines.

\subsubsection{\cii\ 1334 \AA\ and 1335 \AA\ line widths}\label{sec:results_cii_line_widths}

We measured the line widths of \cii\ 1334 \AA\ and 1335 \AA\, where the line width for each given line was defined as 
$\text{w}_\text{FWHM} = 2\sqrt{2\ln 2}\sigma$, where $\sigma$ is the standard deviation of a Gaussian, and thus $\text{W}_\text{FWHM}$ is the full width at half maximum of the same Gaussian. This follows the approach outlined in 
\cite{2015ApJ...811...80R},
where it is shown through the use of synthetic profiles and a model atmosphere that the use of Gaussian fits of \cii\ 1334 \AA\ and 1335 \AA\ line profiles yield useful estimates for $\text{w}_\text{FWHM}$ using the Gaussian standard deviations, even in the case that the two lines are formed under optically thick conditions and when they exhibit double peaks.
The measured $\text{w}_\text{FWHM}$ line widths for \cii\ 1334 \AA\ and 1335 \AA\ for datasets A and B are given in Table \ref{tab:PMJ_cii_stats}, together with their standard errors. 

The line widths of the \cii\ 1334 \AA\ and 1335 \AA\ lines were shown to be correlated to the non-thermal velocity at the heights at which the cores of the two respective lines are formed in 
\cite{2015ApJ...811...81R}. 
The line width is also determined by the thermal velocity, but the line width was further shown to be predominately determined by the non-thermal velocity for line widths greater than $6$ \kms. Profiles with widths smaller than $6$ \kms\ were shown to have dominant optically thin components. In Sect. \ref{sec:results_cii_integrated_intensity_ratio}, we show that for the case of our observed PMJs, the line ratios, $\text{R}_\text{I}$, imply optically thick line formation. For both dates of our observations and for those PMJ profiles in the \cii\ lines that exhibit emission, the line widths exhibit values greater than $6$ \kms. As such, our measured line widths can be used as a diagnostic to probe the non-thermal velocity at the formation heights of the \cii\ line pair. 

To estimate the non-thermal line widths, we used the relation 
\beaq
\text{w}_\text{nth} &=& \sqrt{\text{w}_\text{FWHM}^2 + \text{w}_\text{th}^2 + \text{w}_\text{I}^2}.
\eeaq 
Here, $\text{w}_\text{nth}$ is the estimate of the non-thermal line width for any given observed value of $\text{w}_\text{FWHM}$ (given in Table \ref{tab:PMJ_cii_stats}), $\text{w}_\text{th}$ is the thermal line width, and $\text{w}_\text{I}$ is the instrumental line width of IRIS in the far ultraviolet (FUV) bandwidth.
The instrumental width of IRIS in the FUV is $\text{w}_\text{I} = 12.8$ m\AA\
\citep{2014SoPh..289.2733D}, or $\text{w}_\text{I} = 5.84$ \kms\ for both \cii\ 1334 \AA\ and 1335 \AA.

In order to calculate estimates for $\text{w}_\text{nth}$, we first computed a fitting theoretical estimate for $\text{w}_\text{th}$, for which we used the expression
\beaq 
\text{w}_\text{th} &=& \sqrt{\frac{8\ln(2)k_B T_\text{ion}}{m_\text{ion}}}.
\eeaq
Here, $k_B$ is the Boltzmann constant, $T_\text{ion}$ is the temperature of the \cii\ ion, and $m_\text{ion}$ its mass.

In \cite{2015ApJ...811...81R,2015ApJ...814...70R} 
it was found that the \cii\ lines are formed mainly in the optically 
thick regime, as we observe, and with a mean formation temperature of 
$10$ kK. More specifically, in 
\cite{2015ApJ...811...81R} 
it was found that when the source function for the \cii\ lines has a 
peak in the low atmosphere, this leads to steep emission flanks and a double peak intensity profile. They show an example of this (their Fig. 17) for which the temperature in the line forming region is 
$9.6$ kK. Since we see a majority of double peak profiles and as other results already indicate that PMJs are due to heating events in the lower atmosphere, a temperature of $T_\text{ion} = 10$ kK is thus a reasonable estimate for our purposes, and we find a corresponding thermal line width for the \cii\ ion of $\text{w}_\text{th} =  6.19$ \kms.

Finally, we thus find mean non-thermal line width values of $\text{w}_\text{nth, 1334, A} = 30.2$ \kms\ and $\text{w}_\text{nth, 1334, B} = 42.3$ \kms\ for the 1334 \AA\ line for datasets A and B, respectively, and  $\text{w}_\text{nth, 1335, A} = 40.8$ \kms\ and $\text{w}_\text{nth, 1335, B} = 50.7$ \kms\ for the 1335 \AA\ line for datasets A and B, respectively. We refrain from providing uncertainty values for these non-thermal line widths, as the assumed formation temperature of $T_\text{ion}$ alone would render these dubious at best. As we can see, we find a range of values for the two lines and datasets of $\text{w}_\text{nth} \approx (30,51)$ \kms. The non-thermal line widths are also affected by opacity broadening, which most likely skews them to higher values as typical opacity broadening values are of the order of 1.2-4 \citep{2015ApJ...811...81R} due to optically thick line formation and the double peak profiles that are broader than single peak profiles. Thus, our inferred non-thermal line width ranges serve as upper limits for non-thermal velocities in the formation region of the \cii\ line pair at the upper chromosphere or just below the TR.

\begin{table*}[!ht]
  \begin{center}
\begin{threeparttable}
  \caption{Average values together with their standard errors for different line diagnostics of the \siiv\ line pair for the detected PMJs for datasets A and B are given.}
\begin{tabular}{l l l l} 
\hline
\\
Dataset &  \siiv\ 1394 \AA\ core-offset [\kms]& \siiv\ 1403 \AA\ core-offset [\kms]  & $\text{R}_\text{core} = \frac{\text{I}_\text{1394}}{\text{I}_\text{1403}}$ [ratio] \\
\hline
\\
A (N=33)& $  0.04\pm 0.77$ {\small (N=26)} &$ 1.07\pm 1.19$ {\small (N=17)} &      $ 2.00\pm 0.09$ {\small (N=29)} \\
\hline
\\
B (N=44)& $  0.89\pm 0.84$ {\small (N=35)} &$ 2.09\pm 1.11$ {\small (N=27)} &      $ 1.92\pm 0.07$ {\small (N=43)} \\
\hline
\end{tabular}
\label{tab:PMJ_siiv_stats}
\end{threeparttable}
\end{center}
\end{table*}
\subsection{PMJs in the \siiv\ line pair}\label{sec:results_siiv}

The \siiv\ 1394 \AA\ and 1403 \AA\ lines provide useful diagnostics to probe the appearance and behaviour of PMJs in the TR. We have made use of two specific diagnostics in these lines, namely their respective line core Doppler offsets (Sect. \ref{sec:results_siv_line_core_offsets}) and the line pair intensity ratio of the two lines as well as their basic profile shape (Sect. \ref{sec:results_siiv_line_pair_ratios}). The measured diagnostics for \siiv\ line pair for datasets A and B are given in Table \ref{tab:PMJ_siiv_stats}, together with their standard errors. 

\subsubsection{\siiv\ 1394 \AA\ and 1403 \AA\ line core offsets}\label{sec:results_siv_line_core_offsets}

The line core offsets of the \siiv\ 1339 \AA\ and 1403 \AA\ are proxies for the line of sight velocity at the TR formation height of their cores. We ascertained the line core offsets of both lines by fitting PMJ \siiv\ line profiles with a Gaussian profile and using its centre as an estimate. The means of these measured line core offsets for the \siiv\ line pair with their standard errors and sample sizes are given in Table \ref{tab:PMJ_siiv_stats} for both datasets.

The sampling sizes for both dates and both lines are indicative of the number of profiles in each line for which a Gaussian profile could be fitted with a Gaussian $\chi^2$ goodness-of-fit value (using the IDL `gaussfit' function) that proved suitable to select those profiles with large enough intensity enhancements and with profiles that are well-shaped enough to extract meaningful line core offsets. For peak-brightness times, these profiles were then verified by eye. These sample sizes therefore also present the number of profiles in each case that correspond to events with visible enhancement in the given \siiv\ line for the given date.

From Table \ref{tab:PMJ_siiv_stats} we see that the mean line core offsets for both dates and both lines are consistently of positive values, and they are close to zero. They have a maximum absolute value of $\le 3.20$ \kms, taking standard errors into consideration. As such, we find that the line of sight velocity at the site of PMJ signals in the \siiv\ line pair at the shared approximate height of line core formation for the two lines is likely no more than a few \kms\ and may be close to zero given our standard errors.

\subsubsection{\siiv\ 1394 \AA\ and 1403 \AA\ line core intensity ratios and line morphology}\label{sec:results_siiv_line_pair_ratios}

Our \siiv\ line pair profiles exhibit single peaks and are clearly Gaussian in shape for those cases in which we observe any appreciable intensity enhancement. The line core intensity ratios of the \siiv\ 1394 \AA\ and 1403 \AA\ lines can be utilised to estimate whether the lines form under optically thick or thin conditions. We calculated the peak intensity ratios of the \siiv\ 1394 \AA\ and 1403 \AA\ lines, which are defined such that the ratio is given by $\text{R}_\text{core} = \frac{\text{I}_\text{1394}}{\text{I}_\text{1403}}$. Here, $\text{I}_\text{1394}$ and $\text{I}_\text{1403}$ are the instrument-count intensities at the position of the line core of the given line. 

Under optically thin conditions, the line core intensity ratio for the line pair has a value of $\text{R}_\text{core} = 2$, owing to the fact that the 1394 \AA\ line has an oscillator strength that is twice that of the 1403 \AA\ line. If the lines form under optically thin conditions, the 1394 \AA\ should therefore also present twice the absolute intensity in the line core as that of the 1403 \AA\ line. Nonetheless, line formation under optically thick conditions can also produce a line ratio of $\text{R}_\text{core} = 2$, and thus this value does not guarantee optically thin formation, but a significantly lower value than this does thus preclude thin formation.

We find typical values for $\text{R}_\text{core} \approx 2$. The average intensity line core ratio of the two lines is close to two for both datasets, as shown in Table \ref{tab:PMJ_siiv_stats}. The range of line core intensity ratios over both dates is $(1.85 - 2.09)$, taking standard errors into account. We can thus conclude that the \siiv\ 1394 \AA\ and 1403 \AA\ line PMJ profiles may form under optically thin conditions.

\subsection{A PMJ in the \oiv\ line}\label{sec:results_oiv}

We report a single instance of a PMJ exhibiting a clear intensity enhancement in the \oiv\ line. This line is more commonly associated with flaring activity in the TR and has a formation temperature of around $140,000$ K under equilibrium conditions. It is a diagnostic for lower density plasma than the Si IV lines and as such, it is remarkable that in at least one instance one of our PMJ events appears in this spectral line. The PMJ event in question is dubbed PMJ B21. This event is also our most pronounced PMJ event throughout all of our diagnostics, and it exhibits particularly strong intensity enhancements in most of our studied spectral lines, and especially so in the Si IV line pair. It clearly stands out in \ref{fig:30th_all_profiles}, and we see a single \oiv\ profile that is clearly enhanced compared to those of other events but with a rather irregular shape.

We note that the absence of emission in the \oiv\ line for the remainder of our observed PMJs can be interpreted to be due to the fact that PMJs are mostly a phenomenon in the high density, deep solar atmosphere. The formation of the O IV lines is affected by non-equilibrium ionisation \citep{2013ApJ...767...43O,2016ApJ...817...46M}.

\subsection{Concurrent PMJ darkenings in \halpha\ and \efft\ and their connection to brightenings in \mgii.}\label{sec:results_halpha}

We present observations of darkenings in the inner wings of the \halpha\ line associated with PMJ brightenings as well as darkenings in the inner wings of the \efft\ line. Secondly, we present the spatial relationship between the darkenings in both of these lines and the PMJ brightening we observe in \mgii\ SJ images.

\subsubsection{PMJ darkenings in \halpha\ and \efft.}\label{sec:results_halpha_8542_overlap}

%======================================= fig 9
\begin{figure}[!t]
\centering
\includegraphics[width=8.0cm]{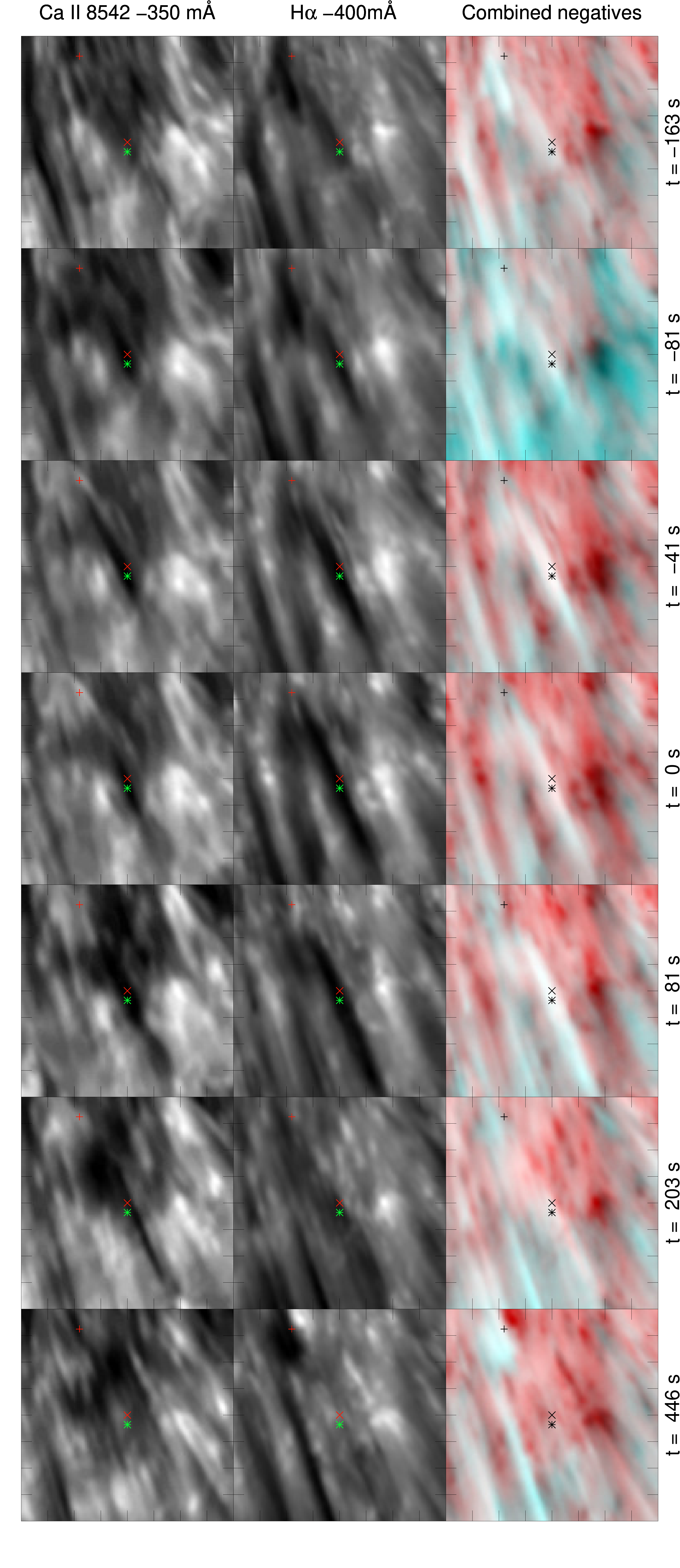} 
\caption{\label{fig:time_evolution_darkening_example_A10} 
PMJ A10 and associated dark features in CRISP images in the \efft\ line at offset $-350$~m\AA\ (left column), the \halpha\ line at an offset of $-400$~m\AA\ (middle column), and a value-inverted combination of these (right column), which is shown at indicated time offsets (time increasing top to bottom). Tick-marks are spaced 1'' apart. Time $t = 0$~s is when the PMJ achieves maximum brightness at \efft\ $-350$~m\AA\ (marked with a plus sign). The combined image displays (re-scaled) negatives of the two CRISP images at their respective offsets, with bright red indicating dark \efft\ $-350$~m\AA\ features and bright cyan indicating dark \halpha\ $-400$~m\AA\ features, when dark features overlap in the two channels, they combine to white in the combined image. Sample positions used for sampling profiles are indicated as follows: plus sign, \efft\ line (bright feature); cross, \efft\ line (dark feature); and asterisk, \halpha\ line (dark feature). }
\end{figure}
%=======================================

%======================================= fig 10
\begin{figure*}[!th]
\centering
\includegraphics[width=18.2cm]{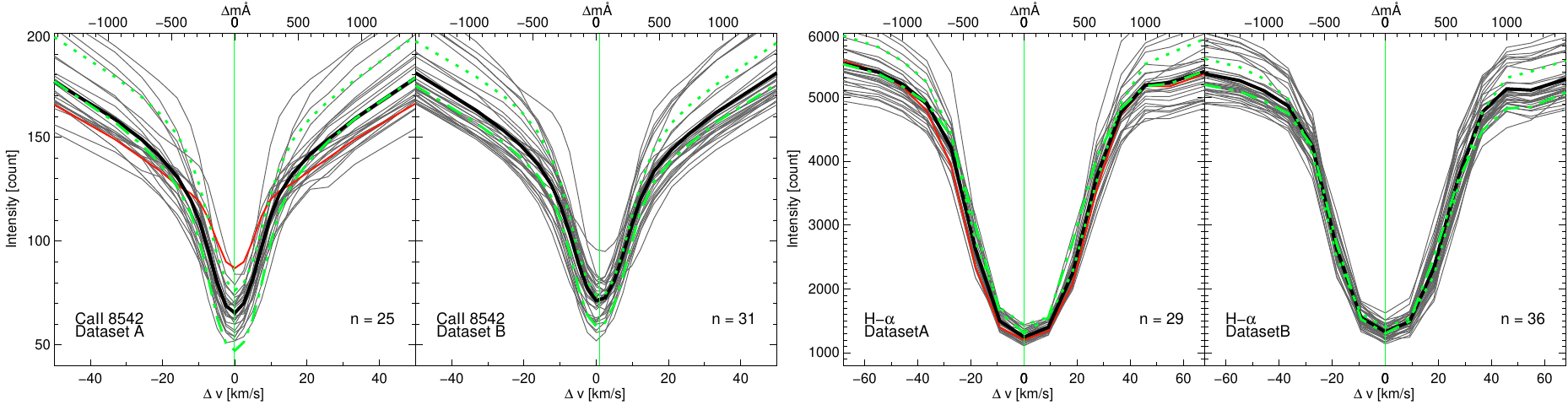} 
\caption{\label{fig:8542_halpha_dark_profiles} 
Spectral signatures of PMJ dark features and reference average profiles in the \efft\ and \halpha\ line from the CRISP instrument for datasets A and B as indicated. The different line styles for the various \halpha\ line profiles denote the following: grey solid lines, the line profiles of dark features for PMJs that clearly exhibit such features in the observations; red solid line, spectral profiles of event A10; black solid line, {the average profile of this sample of PMJ dark features;} green-dotted line, the average line profile across all pixels in the observations; and green-dash-dotted line, the average line profile in the penumbra of AR12533.}
\end{figure*}
%=======================================

%======================================= fig 11
\begin{figure}[!ht]
\centering
\includegraphics[width=8.2cm]{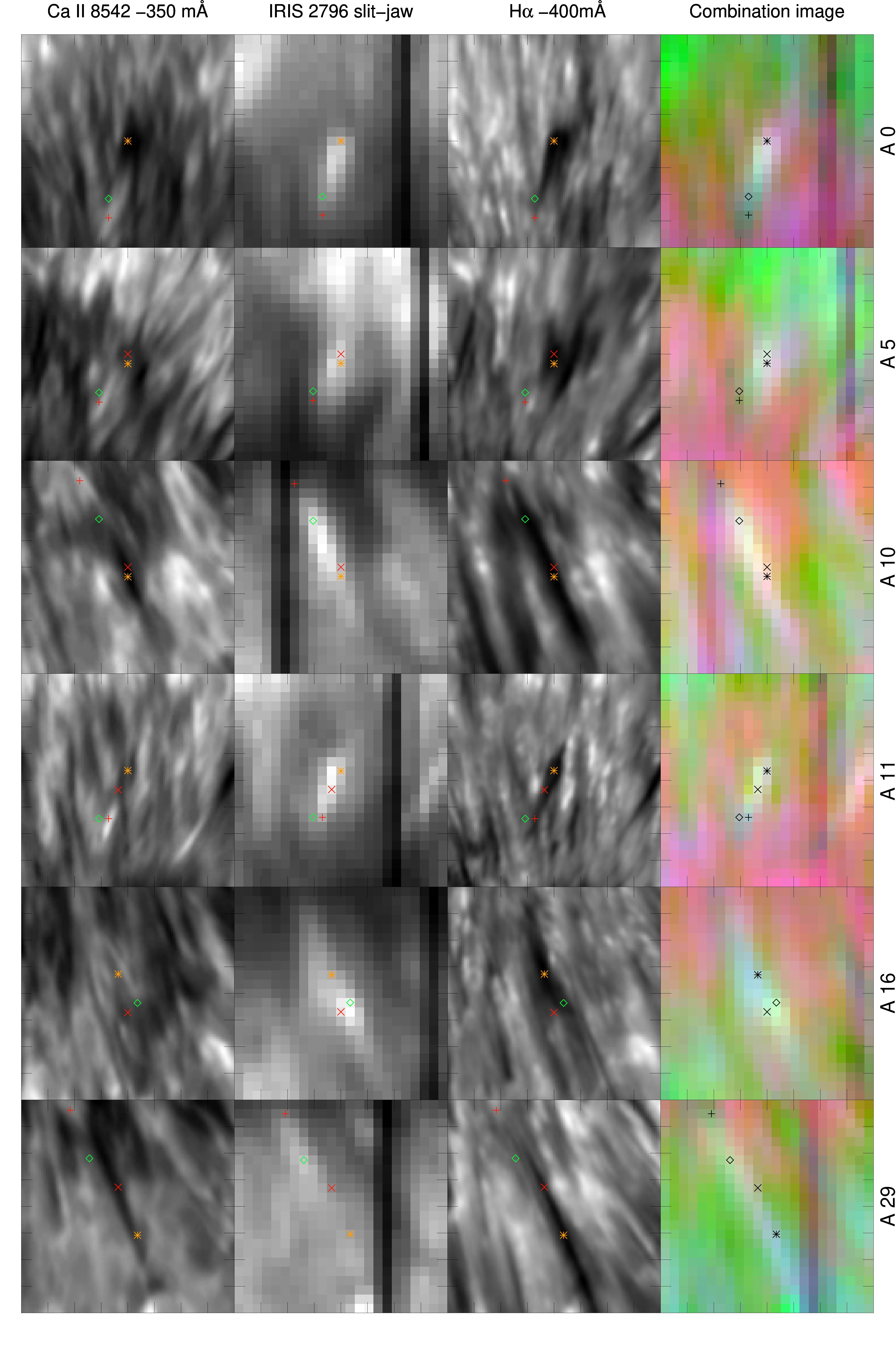} 
\caption{\label{fig:dark_examples_29th} 
PMJs drawn from dataset A for which dark features in the \efft\ blue wing (first column) and the \halpha\ inner blue wing (third column) overlap with brightening in the \mgii\ SJI channel (second column). Tick marks are spaced 1'' apart. PMJs are shown at nominal peak brightness across channels, and each row corresponds to one PMJ with images across channels at the nearest possible timesteps for different instruments, with PMJ identifications indicated for each row on the right. An RGB image combines the channels (fourth row); the intensity inverted channels (bright to dark) of the \efft\ blue wing (red), inner \halpha\ blue wing (blue), and the \mgii\ SJI image (green) are shown. }
\end{figure}
%=======================================

PMJs have mostly been studied in the \caii\ lines, and not so much in the \halpha\ line. Recently, however, 
\cite{2019ApJ...876...47B} 
reported on PMJs that present with darkenings in \halpha\ images when studied by eye and are viewed in the inner line core, with accompanying \halpha\ line profiles exhibiting subtle decreases in intensity in the inner line core at sites adjacent to PMJs that are visible in concurrent \efft\ line observations.
We investigated the appearance of PMJs in \halpha\ and \efft\ images together with their line profiles at selected locations. We confirm the appearance of dark features in the inner wings of the \halpha\ line associated with PMJs. 

We also report the existence of dark features in the inner wings of the \efft\ line that strongly correlate spatially and temporally with those in the inner line wing of the \halpha\ line. Utilising CRISP images at line core offsets of $-400$~m\AA\ and $-350$~m\AA\ in the \halpha\ and \efft\ lines, respectively, we found elongated dark features that trace the direction of the surrounding penumbral fibril structures and that sit at the origin of PMJs, preceding and following PMJs in time. We selected sampling positions by eye, targeting dark features for the subset of PMJs for which we could positively identify them.

Figure \ref{fig:time_evolution_darkening_example_A10} shows the temporal evolution of example PMJ A10. The PMJ is a very clear example of a PMJ that presents with typical brightening in the \efft\ line at an offset of $-350$~m\AA\, but that also exhibits dark features in both the \halpha\ and \efft\ inner line wings, which both have their approximate footpoints at the site where the PMJ bright feature appears. 

In the composite images, bright red indicates the presence of dark \efft\ $-350$~m\AA\ features and bright cyan indicates dark \halpha\ $-400$~m\AA\ features. Co-spatial dark features in the two channels appear white when they are combined in their respective composite images. 
The dark features both precede and follow the bright feature in the \efft\ line by over 2 minutes. In the case of the dark feature visible in the \efft\ line, it becomes rather obscured at peak enhancement of the \efft\ bright-feature, and thus it is less identifiable at this time, which probably made identification less likely in earlier observations and studies.

PMJ A10 is a particularly clear example of the behaviour described above. However, we find numerous PMJs in both datasets for which the same trend of near-cospatial dark features in the inner line wings of the \halpha\ and \efft\ lines, both of which seem to originate at the approximate point of the \efft\ bright feature, holds true. Typically the dark features also precede and follow the bright feature in time in the other cases. An additional example of this type is included in the Appendix in Fig. \ref{fig:time_evolution_darkening_example_B21} for the PMJ B21.

We selected all PMJ events for which we could detect dark features in either or both of the inner line wings of the \halpha\ and \efft\ lines and determined their spectral profiles in these lines at the time of maximum brightness for the typical bright feature that is visible in the \efft\ line. For observations in the \efft\ line, we selected a number of clear sampling positions and dark features of $\text{N}_\text{dark}\left(\text{\ion{Ca}{ii}, A} \right) = 25$ (= 76\%) and $\text{N}_\text{dark} \left( \text{\ion{Ca}{ii}, B} \right) = 31$ (= 70\%) for datasets A and B, respectively. For the \halpha\ line we found a number of equivalent dark features of $\text{N}_\text{dark} \left( \text{\halpha, A} \right) = 29$ (= 88\%) and $\text{N}_\text{dark}\left( \text{\halpha, B} \right) = 36$ (= 82\%) for datasets A and B, respectively. The percentages are all in relation to the total number of PMJs detected for each dataset. We see that a significant majority of PMJs in both lines exhibit these dark features; there are fewer of them for observations in \efft. This is most likely both due to the \efft\ PMJ bright features obscuring some of the \efft\ dark features, as well as a general lower contrast in the \efft\ observations compared to those in the \halpha\ line.
Figure \ref{fig:8542_halpha_dark_profiles} shows all individual \efft\ and \halpha\ PMJ profiles for these dark features, together with the PMJ mean profiles of these. 

The average PMJ profiles in the \halpha\ line for both dates confirm a slight decrease in intensity in the inner line wings of the studied PMJs, which is consistent with the profiles presented in 
\cite{2019ApJ...876...47B}. 
The average \efft\ line profiles for our selected dark features show less distinct behaviour, and they have mean profiles that are overall slightly enhanced compared to the penumbral mean profile, even though the CRISP images at an offset of $-350$~m\AA\ clearly show features that are darker than their surroundings. 
The darkenings found in the images of the inner line wing of the \halpha\ line appear very similar to those as presented in 
\cite{2019ApJ...876...47B}. 
The spectral formation process of the dark features in the inner ling wings of both the \efft\ and \halpha\ lines has yet to be explained and warrants further investigation, possibly in the form of modelling.

Given that PMJs may be caused by reconnection events and the subsequent heating that takes place in the low chromosphere or photosphere, one may expect to find enhancement in the outer wings of the \halpha\ line as observed for Ellerman bombs (see for example
\cite{2013JPhCS.440a2007R}). 
 So far however, we have not found evidence of any such enhancement in \halpha\ profiles for PMJs. This can be seen in Fig. \ref{fig:8542_halpha_dark_profiles} for the sampling positions of the inner wing of the \halpha\ line dark features. The lack of any consistent enhancement in the outer wings of \halpha\ compared to the penumbral and full FOV line profile average may warrant further investigation in the future, especially in light of evidence for heating at chromospheric or photospheric heights, such as enhancement in the \mgii\ triplet blend.  
We conclude that dark structures that precede and follow the nominally bright PMJ events in time exist in both the inner line wing of the \halpha\ line as previously reported, but also in the inner line wing of the \efft\ line.

\subsubsection{Dark PMJ features in \halpha\ and \efft\ and PMJ brightenings in \mgii}\label{sec:results_halpha_efft_mgii_overlap}

Figure \ref{fig:dark_examples_29th} shows a selection of PMJs from dataset A that show clear dark features in the blue wing of the \efft\ line, the inner blue wing of the \halpha\ line, and that exhibit clear brightenings in IRIS \mgii\ SJ images, together with an amalgamation RGB image that combines these three channels. The RGB image uses the intensity-inverted SST images for the \efft\ (red) and \halpha\ (blue) channels. As such, darkenings yield brighter colours for these two channels. The concurrent darkenings in the two channels that overlap with brightenings in the \mgii\ SJ images thus tend towards bright (white) values. The figure illustrates a clear trend in which the brightening in the chromospheric \mgii\ SJI images overlaps with the darkenings in the \efft\ and \halpha\ lines. We find that this trend is typical for both datasets of PMJs and may indicate that these darkenings map to the same chromospheric object as the brightening in the \mgii\ channel.
PMJs from dataset B that were selected following the same criteria as described above are shown in Fig. \ref{fig:dark_examples_30th}, which is included in the Appendix. 

\subsection{Potential twisting of PMJs from \mgii\ h \& k}\label{sec:results_mgii_doppler}

\cite{2018ApJ...869..147T} 
present evidence for the presence of twisting motions in large penumbral jets (LPJs) from an analysis of \mgii\ k line profiles. 
\cite{2016ApJ...816...92T} and 
\cite{2018ApJ...869..147T} 
describe LPJs as jets that are, on average, larger than PMJs and that are found predominantly on the edge of the penumbra. We do not make a distinction between LPJs and PMJs in our own work, and many of our PMJs would constitute LPJs under the working definition in  \cite{2016ApJ...816...92T}.
Here we investigate the presence of twisting motions in our samples of PMJs.

\cite{2018ApJ...869..147T} 
used \mgii\ k Dopplergrams to determine to what degree the observed LPJs exhibit rotation. For an LPJ that rotates around a central axis and is intersected by an IRIS raster slit at an angle close to perpendicular, \mgii\ k (or h) Dopplergrams should exhibit a blue shift on one side, a red shift on the opposite side of the LPJ, and little or no shifts at its centre. Conversely, for the case of an LPJ intersecting a raster slit while closely aligned along its length, one would expect the possibility of Dopplergrams in which only a red or a blue shift is visible. It must be noted that red or blue signals in Dopplergrams are not necessarily due to true Doppler shifts in the line profiles, and they may instead be caused by asymmetry in the intensity enhancement of the blue and red peaks.

\cite{2018ApJ...869..147T} 
found a significant number of LPJs that exhibited Doppler shifts going from blue to red (or vice versa) in the \mgii\ k line along the length of the IRIS raster slit (11 total, 65\% of all LPJs), and also a significant number that exhibited only red  or blue shifts (6 total, 35\% of all LPJs). All LPJs were observed to exhibit either twisting or a distinct blue or red shift. 

We utilised both the \mgii\ k and h line in our investigation. We generated Dopplergrams and bisectors for all of our detected PMJs in the two lines, not restricting ourselves to any specific angular alignment between our PMJs and the raster slit which caught the event. If twisting is present in PMJs, there should be the possibility of detection of at least one Doppler signal 
 component, irrespective of the angle between the PMJ and raster slit, given that the PMJ is not intersected perfectly along its length. 

\cite{2018ApJ...869..147T} 
present one LPJ event that exhibited particularly strong evidence for rotational motion in the form of a strong skew of the \mgii\ $\text{k}_2$ peaks first to red and then to blue upwards along the slit. We include this event in our analysis as a reference (labelled C0). The event was observed on 2015 August 5, at 16:01:54 UT (see 
\cite{2018ApJ...869..147T}
for more details).

{We computed Dopplergrams using a simple approach mimicking the one employed in} 
\cite{2018ApJ...869..147T}, with some differences. To create Doppler maps for our PMJs (and C0), we selected spectral positions at offsets of $\pm 40$ \kms\ in the \mgii\ k\&h lines, subtracting the intensity at the (negative) blue offset from that at the (positive) red offset, then we normalised this value.
In 
\cite{2018ApJ...869..147T}, the Dopplergrams were constructed using offsets at $\pm 50$ \kms\ in the \mgii\ k line, and they were then normalised using a constant factor used across all LPJs. In our observations, our PMJ profiles are on average somewhat narrower, and thus we opted for a narrower offset-range. We normalised our intensity differences using an event-dependent value. We found this necessary as a constant normalisation factor was unsuited to the wide variety of intensity values observed for different PMJs. Highly energetic PMJs with high intensities may exhibit deceptively visually striking blue or red signals in Dopplergrams when intensity differences are normalised by a constant factor that is also used for low-intensity events. We opted for a normalisation factor that is dependent on the average peak intensity of a given PMJ \mgii\ k or h line profile. For each PMJ event, we computed the mean intensity of the two peaks present in the given line and subtracted the mean far-wing intensity. We thus obtained an approximate measure of the total intensity range for both profiles for any given event, $\text{I}_\text{peak range}$. We then used the half value of this PMJ- and line- dependent value to normalise the Doppler-intensity differences along all other pixels along all raster slits for a given PMJ. Thus, our normalised intensity Doppler intensity for either \mgii\ line is defined as $\text{I}_\text{doppler} = \left(\text{I}_\text{-40}-\text{I}_\text{+40}\right)/\left(\text{I}_\text{peak-range}/2\right)$. Here, $\text{I}_\text{doppler}$ is the normalised Doppler-intensity difference, and $\text{I}_\text{-40}$ and $\text{I}_\text{40}$ are the intensities at $\pm 40$ \kms and $\text{I}_\text{peak range}$ is the scaling intensity described above. A value of one in the Doppler maps thus corresponds to a relative intensity difference as large as half the mean peak intensity enhancement of the event, and it indicates that the blue wing is more enhanced than the red one. A value of $-1$ indicates the reciprocal.

%======================================= fig 12
\begin{figure}[!t]
\centering
\includegraphics[width=8.5cm]{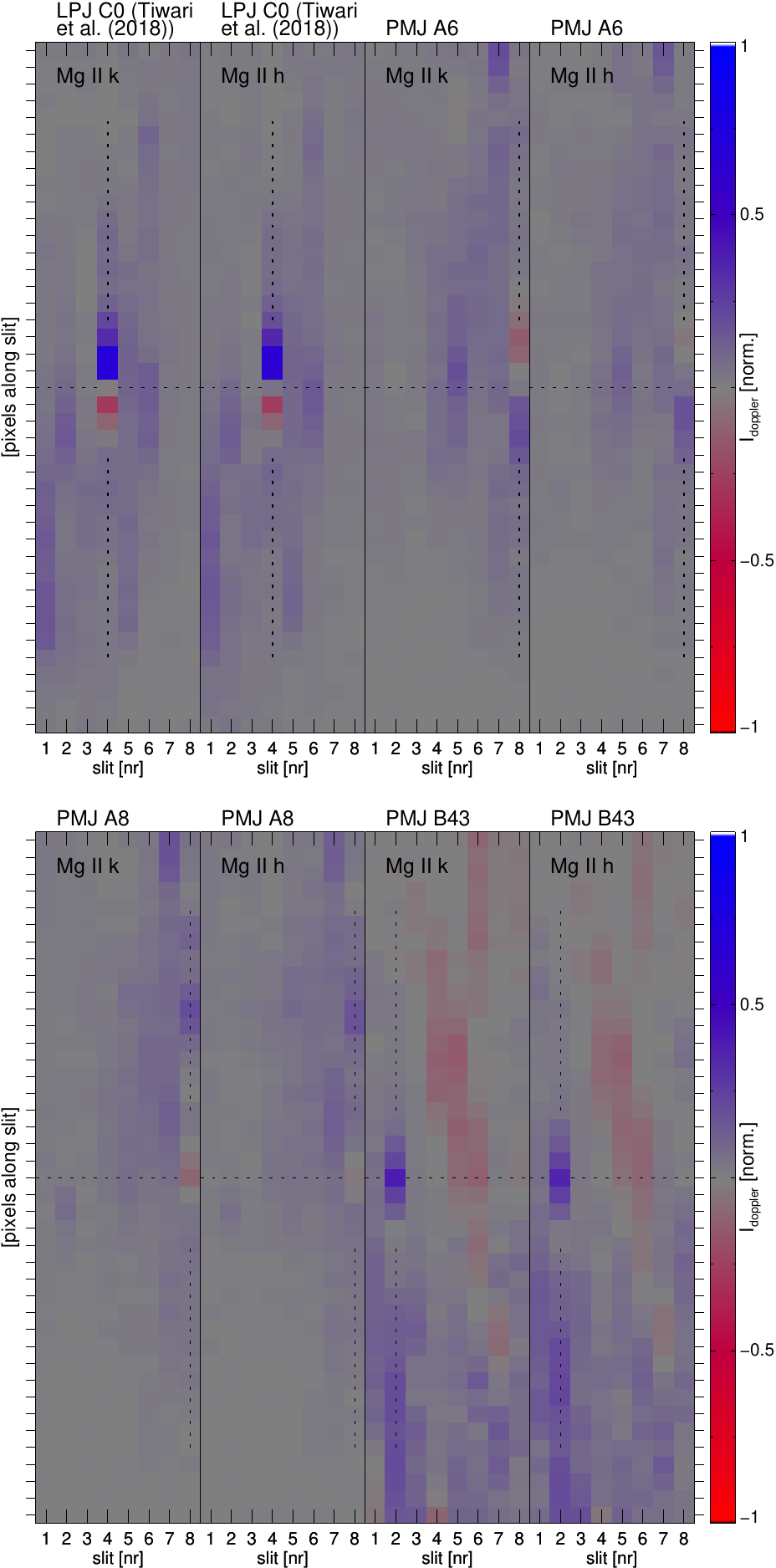} 
\caption{Doppler maps in the \mgii\ k \& h lines for a selection of PMJ and LPJ events. The Doppler maps cover all eight IRIS raster-slit positions along the x-axes and 41 pixels along the y-axis, with the nominal \mgii\ h \& k sampling position at the centre of the y-axis. Horizontal dashed lines enclose the sampling position, and vertical dashed lines surround the sampling position and the three pixels above and below it. Maps are labelled with the relevant \mgii\ line and event identifiers. Doppler map values have a range of $\left(-1,1 \right)$ with negative values in red, signifying red enhancement, and positive values in blue, signifying blue enhancement. The intensity value $I_\text{doppler}$ is defined in the text.\label{fig:pmj_dopplergram_examples} }
\end{figure}
%=======================================

%======================================= fig 13
\begin{figure*}[!t]
\centering
\includegraphics[width=17cm]{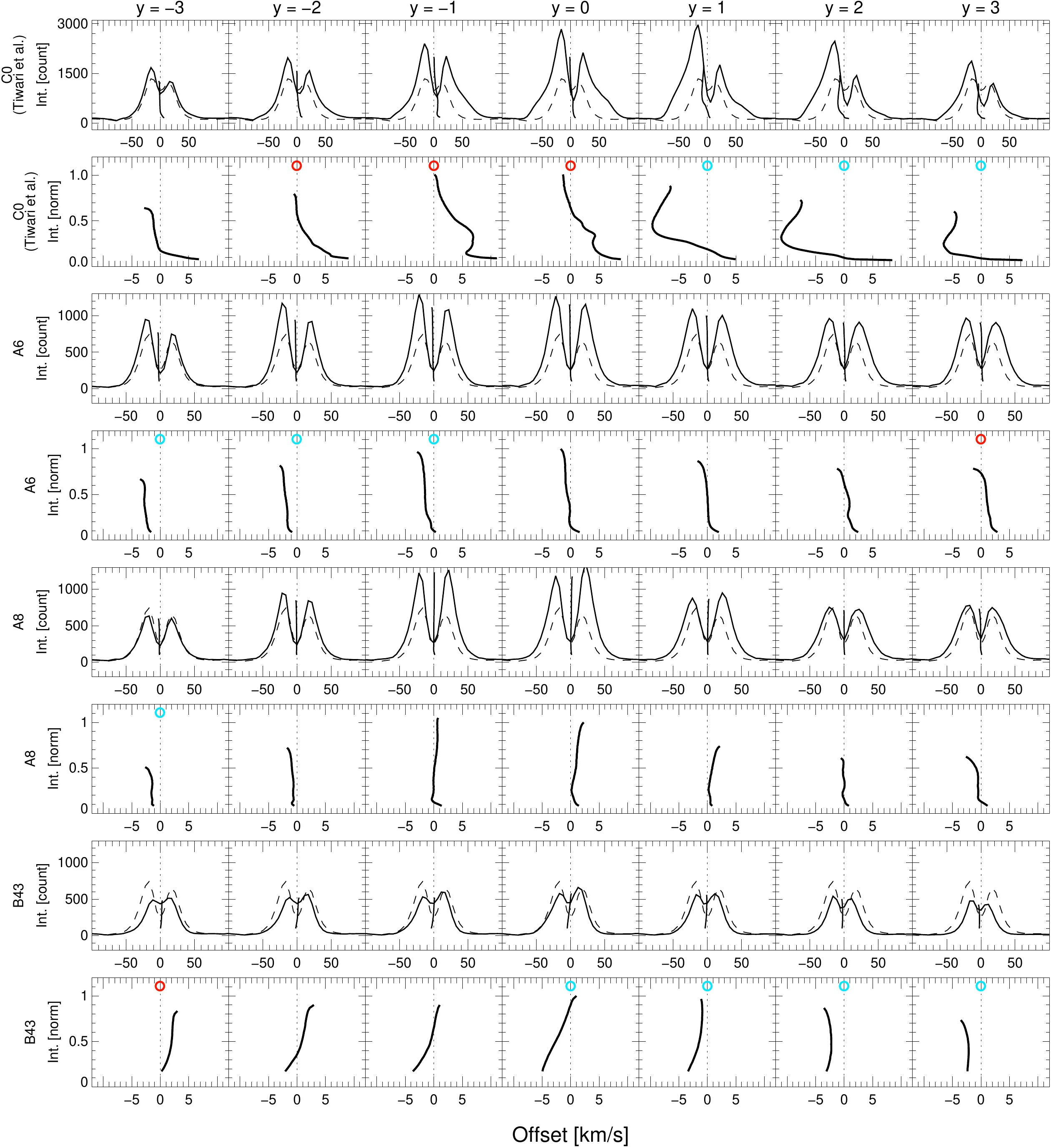} 
\caption{Spectral profiles and their bisectors in the \mgii\ k line for a selection of PMJs and an LPJ. Panels from left to right show spectral profiles and/or bisectors along the relevant IRIS raster slit at y-axis pixel positions from y $= - 3$ to y $= 3$ (as labelled) with y $= 0$ being the \mgii\ h \& k sampling position of the event. For each given y-pixel position, the top row shows for LPJ event C0 (see text for details) both the \mgii\ k spectral profile and its associated bisector (both black solid lines) together with the mean \mgii\ k profile over the entire field of view of event C0's observations (black-dashed line). The second row displays only the \mgii\ k bisectors (black solid line) of the same event; panels for bisectors with mean values > $|\pm 1|$ \kms\ are marked with a blue (negative value) or a red circle (positive value). The panels below show the same plots only for selected PMJ events (as labelled on the left), with spectral profiles and bisectors shown together above rows only displaying bisectors. We note that for the PMJ events, the mean spectral profile (black-dashed line) in the plots is instead the mean PMJ profile of the relevant dataset from which the event is drawn.
\label{fig:pmj_bisector_examples} }
\end{figure*}
%======================================

It is not always easy to visually identify the extent of the blue or red shift from Dopplergrams alone and whether reversals along raster slits in profile-types are significant or not. We therefore also calculated bisectors for the \mgii\ h and k lines. We calculated these for all PMJs along a seven-pixel line along a given IRIS raster slit, centred on our primary sampling position in the lines. The bisectors were computed by calculating the interpolated line-offset values in the given \mgii\ line at specific intensity values along both the red and blue slope of the red and blue peak, respectively, on each side of the central minimum. We then calculated the mean of the red and blue wing offsets at these intensities, yielding the values of the bisectors between the outer red and blue slope of the \mgii\ k \& h profiles. We calculated the bisectors up to the intensity of whichever peak of the \mgii\ profiles was the smallest, and down to an intensity found at an offset of $100$ \kms. This avoids bisectors trailing off horizontally as the spectral profiles trend towards the \mgii\ wings.

For Dopplergrams all scenarios of twisting should be visually identifiable by  colour. When inspecting bisectors, one would expect the overall position of the bisectors to shift from red to blue offsets, or the reciprocal, along the IRIS slit in the case for twisting. In the case of solitary shifting of intensity towards the red or blue wings, we would expect the bisectors to exhibit a consistent shift near the actual \mgii\ sampling position and to be unchanged compared to their typical values for pixel positions that are further distant along the IRIS slit.

We find that PMJs have a tendency to have an enhanced blue signal with mean values of $\text{I}_\text{doppler}$ for dataset A, \mgii\ h, $\text{I}_\text{doppler, A} = 0.13$ , and for \mgii\ k, $\text{I}_\text{doppler, A} = 0.13$. For dataset B, \mgii\ h, $\text{I}_\text{doppler, B} = 0.16, $ and for \mgii\ k, $\text{I}_\text{doppler, B} = 0.15$. This is also reflected visually with most bisector lines falling in the blue, but only at modest offsets of a few \kms. 
Notably, this stands in contrast to the line-core offsets given in Sect. \ref{sec:results_mgii}  and Table \ref{tab:PMJ_mgii_stats}, all of which lie in the red although with values < 1 \kms. This behaviour is reflective of the dominance of the blue peaks in both the \mgii\ k \& h lines, which in a great majority of cases are more intensive than their red counterparts. This skews both the Dopplergrams and the bisectors slightly towards the blue in most cases.
We find only a few cases in which our PMJs exhibit behaviour that is clearly consistent with twisting in which we see a reversal of blue-to-red or vice versa along the IRIS slit direction. There are more occasions of an isolated and significant shift of intensity towards the \mgii\ line pair wings, but these are also not ubiquitous.

Figure \ref{fig:pmj_dopplergram_examples} shows \mgii\ k \& h line Dopplergrams produced as described above along all eight IRIS raster slits for the IRIS observation for events C0, A6, A8, and B43.
Figure \ref{fig:pmj_bisector_examples} shows profiles along the seven pixels centred on the \mgii\ sampling position for the same example PMJs. We show only profiles and bisectors for the \mgii\ k line. Bisectors in the \mgii\ h are very similar to those in the other line due to very similar profile shapes, although the k line usually exhibits a greater signal and clearer bisectors.

As reference, in Fig. \ref{fig:pmj_dopplergram_examples}, we clearly observe event C0 to exhibit a transition from red to blue upwards along the IRIS slit, as first described in 
\cite{2018ApJ...869..147T}. 
We see this in the \mgii\ k line, as first shown in 
\cite{2018ApJ...869..147T}, 
as well as in the \mgii\ h line, first shown here.
As a side effect of our event-dependent intensity normalisation, we also observe a less saturated signal in our Doppler maps for this event. 
Event C0's bisectors shown in Fig. \ref{fig:pmj_bisector_examples} also show a clear trend in their overall position and shapes, indicative of a twisting scenario. Here, we go from a bisector with little Doppler shift in either direction at y = $-3$, to progressively stronger red-shifted bisectors for y = $-2$ and y = $-1$, to a slight decline in red shift at y = $0$, and a strong reversal in the bisectors mean value at y = $1$ towards the blue, with the bisector at y = $2$ being shifted even more towards the blue, and the bisector at y = $3$ again diminishing in overall Doppler signal.  

Event A6 and A8 exhibit the visually clearest examples of Doppler-signal reversals in our Doppler maps in all our observations, and they are both shown in Fig. \ref{fig:pmj_dopplergram_examples}. The events are located at near-identical positions in the upper right of our observed sunspot, which is close to the outer penumbra, and both originate from the same dataset. Event A6 precedes A8 by only 183 s and as such, the events are most likely an example of the repeating behaviour often seen for PMJs and are strongly related. When observed in the inner blue wing of the \efft\ line, both PJMs appear to be of a similar size, of the order of around 1.5'' - 3'', or approximately 1000 - 2000 km. This size is consistent with PMJ sizes, and it sits on the lower end of the LPJs presented in 
\cite{2018ApJ...869..147T}. 
Both events intersect the IRIS raster slit at an approximately $45^\circ$ angle. Both events exhibit a reversal from blue to red upwards along their IRIS slit, with A6 having a stronger signal than A8, and both exhibiting more marked intensity shifts in the wings in the \mgii\ k line than in the h line. Overall, both events exhibit much smaller differences in the shift of their Doppler-intensity difference than C0, as is evident from their appearance. We note that both events are also notable in that they also correspond to the PMJs that exhibit the strongest signal in the \mgii\ triplet positions, although whether the two behaviours are related is not clear nor investigated. 
The bisectors of the two events are also shown in Fig. \ref{fig:pmj_bisector_examples}, where A6 again shows a clearer transition of overall blue-shifted \mgii\ k bisectors that trend towards the red from y $ = -3$ to y $= 3$, and A8 shows a much more modest transition, but follows the same trend. 

Event B43 is our third-strongest source of evidence for twisting in PMJs. In the Doppler maps given in Fig. \ref{fig:pmj_dopplergram_examples}, the event shows little clear evidence for a reversal in intensity difference along the IRIS raster slit, only clearly showing a strong blue intensity difference at and around the PMJ sampling position. In Fig. \ref{fig:pmj_bisector_examples}, however, a clear shift towards the red in the bisector at y$ = -3$ can be observed, shifting gradually towards blue-shifted bisectors. In this case, the Doppler map most likely fails to indicate this behaviour due to the overall narrowness of the profile, which therefore does not exhibit strong intensities at offsets $\pm 40$ \kms, and so it fails to depict the shift of intensity towards the wings. This serves to demonstrate the diagnostic value of bisectors in detecting such a signal given the variety of intensity ranges in the \mgii\ line pair. 
We detect only two additional events that may show evidence of a transition from blue-to-red or vice versa along the IRIS slit. 

\begin{table}[!t]
  \begin{center}
\begin{threeparttable}
  \caption{Number of occurrences of twisting and presence of enhanced blue or red signal for PMJs in dataset A and B.}
  \begin{tabular}{l l l l}
 Dataset(s) & Twisting & Enhanced & Enhanced \\
 & [N] & Red [N] & Blue [N] \\
 \hline
  A \small{(N = 33)} & 2 (6\%) & 1 (3\%) & 8 (24\%) \\
  B \small{(N = 44)} & 3 (7\%) & 6 (14\%) & 13 (30\%) \\
  A\&B \small{(N = 77)} & 5 (6\%) & 7 (9\%) & 21 (27\%) \\
\hline
\end{tabular}
\label{tab:dopplergram_stats}
\end{threeparttable}
\end{center}
\end{table}

Furthermore, while there are cases of PMJs exhibiting significant intensity enhancement 
in the \mgii\ line pair profiles towards either the red or the blue, they are not ubiquitous. We summarise the number of PMJs that exhibit possible evidence for twisting and those for PMJs that exhibit only increased enhancement towards the red or blue in either of the lines in  
Table \ref{tab:dopplergram_stats}. We find that for the datasets combined that there are 6\% or fewer of the PMJs that exhibit evidence of twisting, and 9\% and 27\% that exhibit a significant red  or blue shift, respectively.
Again, we see a bias in a more numerous occurrence of a blue-ward skew in the profiles.

While we are able to detect the profile shape variation in the \mgii\ line pair along the IRIS raster slit that would be indicative of twisting PMJs, given our discussed diagnostic methods and the few examples we do find, we conclude that there is only a small minority of events that in fact exhibit such behaviour. 
In the supplementary event-summary videos that depict all PMJs in dataset A and B through time, we include plots of bisectors for both the \mgii\ k and h line together with corresponding Dopplergrams for these lines.

We caution against a quantitative interpretation of our Dopplergram signals and bisector-shift values. The shifting of enhancement towards either the red or blue wings of the \mgii\ lines measured by these two diagnostics should not be interpreted as actual velocities. Instead they serve as useful indicators of when line profiles exhibit appreciable asymmetries towards the blue or red in the relevant line and away from the nominally undisturbed average Dopplergram or bisector values. Such a shift in enhancement is consistent with what one would expect for mass velocities in the same direction, but specific values for Doppler velocities should not be inferred from our diagnostics.  

\subsection{On the time evolution of PMJs observed in multiple channels}\label{sec:time_evolution}

We perform a qualitative study of the temporal behaviour of the spectral line samples that are available from our studied PMJs. The available cadence of the IRIS 
spectrograph line profiles we have obtained for select positions of our studied PMJs and which limits our investigation is $\sim40$~s. 

\begin{figure}[!th]
\centering
\includegraphics[width= 7.5cm]{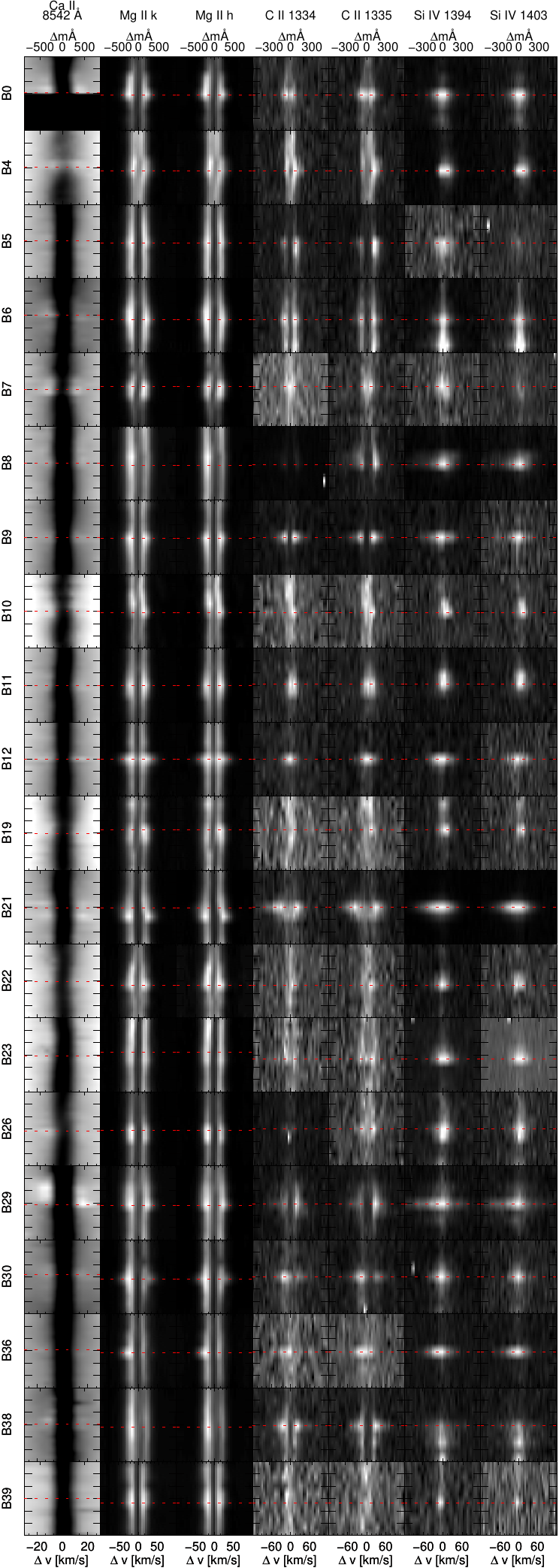} 
\caption{\label{fig:timeslices_30th} Spectral timeslices for a selection of PMJ events from dataset B that display clearly visible signals. Each horizontal row corresponds to a PMJ, marked with its designator to its left. Each column from left to right is labelled with the appropriate spectral line at its top. Horizontal dashed lines in each panel indicate the time of nominal peak intensity of the given PMJ. Each PMJ event is shown for a time of 6 minutes.}
\end{figure}

Figure \ref{fig:timeslices_30th} highlights a selection of PMJs from dataset B, showing timeslices with wavelength along the x-axis and time along the y-axis for the \efft\ line, the \mgii\ h \& k lines, the \cii\ 1334 \AA\ and 1335 \AA\ lines, and the \siiv\ 1394 \AA\ and 1403 \AA\ lines. 
For both datasets, we selected a subset of events with the criteria that they exhibit easily identifiable signals in both the chromospheric and TR spectral lines, and that they do so with clear peaks in intensity in their timeslices. We found the same general appearance and trends in both dataset A and B. For reference, the timeslices for dataset A are included in the Appendix in Fig. \ref{fig:timeslices_29th}.

In inspecting the timeslices of PMJ events, there does not seem to exist a clear trend of spectral signals preceding each other. In particular, we do not observe a temporal trend that correlates with what atmospheric temperature that the channels are usually associated with. This finding is congruent with a scenario in which PMJs light up across their entire length through multiple channels in a relatively short amount of time, rather than becoming apparent at different times in different channels during their lifetime.

\cite{2017ApJ...835L..19S} 
find that PMJs and TR bright dots spatially correlate and that bright dots precede PMJs in time. The authors suggest that bright dots may map to the site of magnetic reconnection at a TR temperature and manifest as PMJs further down in the atmosphere. In the study, half of identified bright dot events identified in IRIS SJI observations are identified to correlate with PMJs that are visible in SOT \cahline\ images and IRIS SJI 2796 observations, thus yielding a sample of 90 bright dot-PMJ linked events. In a large majority of these, the bright dot events precede the corresponding PMJ events by an average of 16.0s and in some cases by up to a minute.

\cite{2017ApJ...835L..19S} 
propose a scenario in which bright dots are caused by reconnection events at TR temperatures, which in turn trigger PMJs at lower chromospheric heights. This would imply that our own PMJ signals in TR-sensitive spectral lines should precede PMJ signals in chromospheric spectral lines in at least some cases, and that the inverse should be true in fewer cases. In our own observations, we would expect to observe this effect despite the slow cadence of our IRIS spectrograph spectral line profile observations of $\sim 40$~s as we sample PMJs close to peak enhancements and since
\cite{2017ApJ...835L..19S} 
report that bright dots usually precede PMJs of the order of 10 s to 1 min. As noted, we do not find a noticeable effect of this nature in our observations.

%======================================= fig 14
%=======================================

\section{Discussion: PMJ origins and evolution \label{sec:discussion}}

In Sect. \ref{sec:results_halpha} we show that PMJs are associated with dark features, not only in the inner line wing of the \halpha\ line as first shown in 
\cite{2019ApJ...876...47B}, 
but also in the inner line wing of the \efft\ line. We also found that the `true' PMJ comprised of the bright jet-like feature in the \efft\ line may, to varying degrees, obscure its accompanying dark feature in the line at peak brightness and through its evolution, the jet-like structure can be seen to extend along the direction of the dark feature. 
The dark features in both lines are well-aligned with the underlying and surrounding fibril structure, and they align to a large degree with the chromospheric brightenings associated with PMJs in \mgii\ IRIS SJ images. The dark features in both lines are also typically present both before and after the main brightening event of the order of seconds to minutes. 
From this behaviour, it is reasonable to assume that the PMJ brightenings in fact develop along a pre-existing fibril structure. As such, this would be congruous with a scenario in which PMJs evolve out of a fibril that undergoes heating due to a reconnection event and lights up along its length as a heating front passes through and along the fibril with little mass motion actually taking place.

This scenario is also supported by a qualitative summary of the analysis of the spectral diagnostics, as laid out in Sections \ref{sec:results_8542} through \ref{sec:results_siiv} in which PMJ responses in the \efft, \mgii\ h \& k lines and the \cii\ and \siiv\ line pairs are investigated. Here, we have found that the values of Doppler offsets and signals 
are modest or close to zero. In fact, taken as a whole, both inferred Doppler-LOS velocities and velocity gradients associated with all studied atmospheric temperatures are modest, and they are much lower than earlier reported apparent PMJ velocities in the literature. This suggests that PMJs do not undergo significant mass motions. 
The measured Doppler offsets for all spectral lines are confounded by the background emission of the chromosphere and TR that surrounds our PMJs. This means that Doppler offsets cannot be interpreted as perfectly analogous to mass flow velocities. However, if PMJs exhibited significant flow velocities, these should still be discernible in our Doppler diagnostics.  

The study of the time evolution of our PMJs throughout different spectral lines in Sect. \ref{sec:time_evolution} shows that the times at which brightenings appear do not differ significantly of the order of our IRIS raster cadence of $\sim 40$~s for the different spectral line sampling positions. This implies that PMJs brighten across all atmospheric temperatures at which they appear concurrently, rather than moving from lower to higher temperatures, for example. As noted in Sect. \ref{sec:time_evolution}, however, the limiting IRIS raster cadence does not permit us to conclude that there is no discrepancy in onset times on timescales smaller than $\sim 40$~s. Nonetheless, we can rule out a systematic discrepancy in PMJ onset times through the different atmospheric channels that points to earlier onset times in channels associated with either low (chromospheric) or hight (TR) temperatures of the order of our cadence. 
Through the use of inversions with the STiC code
\citep{2016ApJ...830L..30D, 2018ascl.soft10014D} 
and other analyses of PMJ observations made using the CRISP and CHROMIS instruments at the SST, 
\cite{2019ApJ...870...88E} 
found that PMJs most likely exhibit only small mass motions. 
In \cite{2017ApJ...849L...7D}, 
advanced MHD simulations are used to explain the high apparent velocities observed for spicules in chromospheric and TR channels. Here, these high apparent velocities are explained by a heating front that moves along the spicule structure and produces a fast apparent brightening, which is not due to high velocity mass motions. 
\cite{2019ApJ...870...88E} 
posit that such a process may also provide a plausible explanation for the formation of PMJs. The authors argue that PMJs may be explained by a propagation of perturbation fronts that originate in the deep photosphere and then dissipate energy within the PMJ. This was supported by findings that place temperature increases in the low chromosphere and with inversions showing that PMJs heat up with increasing height.

The results of 
\cite{2019ApJ...870...88E} 
and our present investigation also fit with the recent results in 
\cite{2019A&A...626A..62R}, 
where fast cadence ($\sim1$ s) \cahline\ observations revealed only modest mass motions for PMJs, but where it was found that PMJs nonetheless light up along their lengths over very short timescales along faint but pre-existing fibrils. The discussion in 
\cite{2019A&A...626A..62R} 
also goes into detail in support of the heating-front PMJ scenario first described in 
\cite{2019ApJ...870...88E}, 
and we would like to direct the reader here. Both of these studies and the fact that they do not find true mass motions are congruent with our present results discussed above, as well as our findings that we see emission for some PMJs in the wings of the \mgii\ triplet blend, indicating heating at upper photospheric heights for at least a subset of PMJs. The latter provides further evidence for a magnetic reconnection event taking place in the deeper atmosphere.

The totality of the results presented here and given the findings discussed above can be interpreted to imply that PMJs are manifestations of heating events most likely originating due to heating events at photospheric to lower chromospheric heights and that propagate upwards as a heating front, with only modest accompanying mass motions. This heating front reaches TR temperatures as evidenced by \siiv\ line pair emission and a single instance of emission in the \oiv\ line. 

As in
\cite{2019A&A...626A..62R}
and
\cite{2019ApJ...870...88E}, 
these conclusions stand in contrast to earlier results in the literature that have observed PMJs to exhibit apparent velocities reaching the order of 100's of \kms. As remarked in greater detail in 
\cite{2019A&A...626A..62R}, 
it is possible that these higher apparent velocity measurements are due to a lack of temporal resolution in the relevant observations. The fast phase of brightening that takes place over the entirety of the PMJs observed in 
\cite{2019A&A...626A..62R} 
can, if observed with slow cadence observations, be mistaken for a fast apparent motion, rather than a fast brightening with no clear directionality.

\section{Conclusions and summary \label{sec:conclusions}}

\subsection{Summary of diagnostics}

\subsubsection{PMJs at different heights in the solar atmosphere}

The different spectral diagnostics that we have used to study PMJs are sensitive to different broad regions in the solar atmosphere. 
In rough order of atmospheric height, from the photosphere and upwards, we can infer the following:

\begin{itemize}
\item Emission in the \mgii\ triplet wings provides evidence for heating at photospheric heights for a subset of PMJs.
\item \mgii\ $\text{h}_2$ \& $\text{k}_2$ average peak Doppler shifts yield estimates for the line of sight velocity in the lower chromosphere that are close to zero.
\item \mgii\ $\text{h}_2$ \& $\text{k}_2$ peak separations are measured to be of the order of 30 \kms. The interpretation of this enhanced separation as a difference in line of sight velocity between the mid- and more upper chromosphere has a high uncertainty. It is more likely a result of enhanced heating in the lower chromosphere.
\item  \cii\ widths yield estimates for the non-thermal line widths and thus upper limits for the non-thermal velocity in the upper chromosphere or just below the TR of $\approx (30,51)$ \kms.
\item \mgii\ $\text{h}_3$ \& $\text{k}_3$ line core offset values yield estimates for the line of sight velocity in the upper cromosphere, with values close to zero.
\item \cii\ line-pair line core offset values yield estimates for the line of sight velocities in the upper chromosphere or TR heights, with absolute values $\le 7$ \kms, ranging slightly below and above zero.
\item \cii\ line-pair ratio values imply optically thick line formation in the upper chromosphere for these lines.
\item The typical number of (double) peaks in the \cii\ line pair implies a peak in the source function of the lines below the upper chromosphere or TR for the majority of PMJs that show an appreciable signal in these lines.
\item Values for the \siiv\ line-pair line core offsets yield estimates for the line of sight velocity at the TR, with modest values of absolute values $\le 3.2$ \kms.
\item The ratio of the \siiv\ line pair peak intensity ratios imply optically thin line formation in the TR for these lines.
\item One example of emission in the \oiv\ line implies heating to TR or flare conditions in one instance.
\end{itemize}

\subsubsection{PMJ dark features in the \halpha\ and \efft\ lines}

PMJs exhibit dark features in the inner wings of the \halpha\ line, which was first shown in 
\cite{2019ApJ...876...47B}. 
We confirmed these \halpha\ line dark features, and further found that PMJs also exhibit dark features in the inner line wing of the \efft\ line, with a large degree of spatial overlap at the two wavelengths. In both channels, the elongated dark features are aligned with the direction of the typical PMJ brightening in the inner line core of the \efft\ line, originating at the head of the jet-like PMJ event. 

At peak brightness, the \efft\ bright feature may obscure the dark feature to some degree. The dark features are observable on timescale of minutes before and after the bright PMJ event itself, and they are aligned with the fibril structure of the sunspot. 

This provides evidence for the scenario in which PMJs are heating events that follow the already existing fibril structure. 
We also find that the PMJ darkenings in the inner wings of the \halpha\ and \efft\ lines are spatially well-aligned with the typical PMJ brightenings in the IRIS \mgii\ SJI channel, indicating that these darkenings are chromospheric events. 

\subsubsection{Twisting motions of PMJs}

By using \mgii\ h \& k PMJ spectral profiles along the length of IRIS raster-slit positions, we produced Dopplergrams and bisector plots in order to investigate potential twisting motions in PMJs. We only found a few events that exhibited a reversal of blue-to-red or red-to-blue enhancement in PMJ \mgii\ h \& k line profiles along IRIS raster slits, which is indicative of twisting motions. Only 5 (6\%) out of the total 77 studied PMJs exhibited Dopplergram or bisector features that are clearly congruent with twisting motions. We found a further 7 (9\%) and 21 (27\%) PMJs that exhibited only red or blue enhancement, respectively, in the \mgii\ h \& k lines. We conclude that twisting motions in the PMJs in our datasets are rare and that they are not a universal property requisite for their formation at the scales we observe.

\subsubsection{Temporal evolution of PMJs}

The temporal evolution of PMJ brightenings when studied across different spectral lines reveals that their onsets do not differ significantly on the timescale of our IRIS raster cadence of $\sim40$~s. As such, we do not observe that PMJs present at a particular temperature or atmospheric layer corresponding to a specific wavelength before others. If PMJs consist of true mass motions, one would expect discrepancies between onset times between different channels. The limiting cadence of $\sim40$~s
for the IRIS raster channel is less than the upper limit of the time range of 10 s to 1 min differences in onset times between bright dots and associated PMJs reported in 
\cite{2017ApJ...835L..19S}.

\subsection{Qualitative summary}

The clear takeaway from this work is that PMJs extend through atmospheric layers from the photosphere to the TR and that they are characterised by low Doppler velocities and velocity gradients throughout all atmospheric layers. Further, PMJs are associated with chromospheric dark features in the inner wings of both the \halpha\ and \efft\ lines and PMJs develop along the direction of these dark features that are present before and after the PMJ brightening in the inner \efft\ wing. The dark features align well with the surrounding fibril structure, which may suggest that PMJs develop along pre-existing fibrils. We find only very few PMJs that exhibit clear indications of twisting. We find evidence of low atmosphere heating in a small number of PMJs due to emission in the wings of the \mgii\ triplet, indicating that some PMJs may result from heating at photospheric heights. The temporal evolution of PMJs through all studied channels indicates no preference for first-onsets in any one channel on the time scale of the $\sim40$~s IRIS raster cadence. 

We find these results to be consistent with the scenario put forth in 
\cite{2019ApJ...870...88E} 
and
\cite{2019A&A...626A..62R} 
that PMJs are the result of a magnetic reconnection heating event at photospheric or low chromospheric heights that propagate heat upwards in the atmosphere through  a heating front that travels along a pre-existing fibril. Furthermore, the heating front then causes a brightening along the fibril at Alfvenic speeds with the PMJ exhibiting only low mass motions. 
A natural next step in elucidating the nature of PMJs would be MHD modelling of PMJs and the production of synthetic spectral profiles constrained by the now extensive spectral diagnostic values and PMJ properties found here, combined with the previous constraints provided by the literature.

\begin{acknowledgements}
The Swedish 1-m Solar Telescope is operated on the island of La Palma 
by the Institute for Solar Physics of Stockholm University in the 
Spanish Observatorio del Roque de los Muchachos of the Instituto de 
Astrof{\'\i}sica de Canarias.
The Institute for Solar Physics is supported by a grant for research 
infrastructures of national importance from the Swedish Research Council 
(registration number 2017-00625).
This research is supported by the Research Council of Norway, project number 250810, and through its Centres of Excellence scheme, project number 262622.
IRIS is a NASA small explorer mission developed and operated by LMSAL with mission operations executed at NASA Ames Research center and major contributions to downlink communications funded by ESA and the Norwegian Space Centre.
This study benefited from discussions during the workshop ``Studying magnetic-field-regulated heating in the solar chromosphere'' (team 399) at the International Space Science Institute (ISSI) in Bern, Switzerland.
We made much use of NASA's Astrophysics Data System Bibliographic
Services.
We wish to thank Tiago Pereira in particular for valuable suggestions and discussions.
\end{acknowledgements}

\bibliographystyle{aa}
\bibliography{biblio}

\begin{appendix}\label{sec:appendix}

\section{Supplementary figures}

Here we provide supplementary figures showing more examples of specific PMJ behaviours as described in the preceding text. 

\begin{figure}[!th]
\centering
\includegraphics[width=8.0cm]{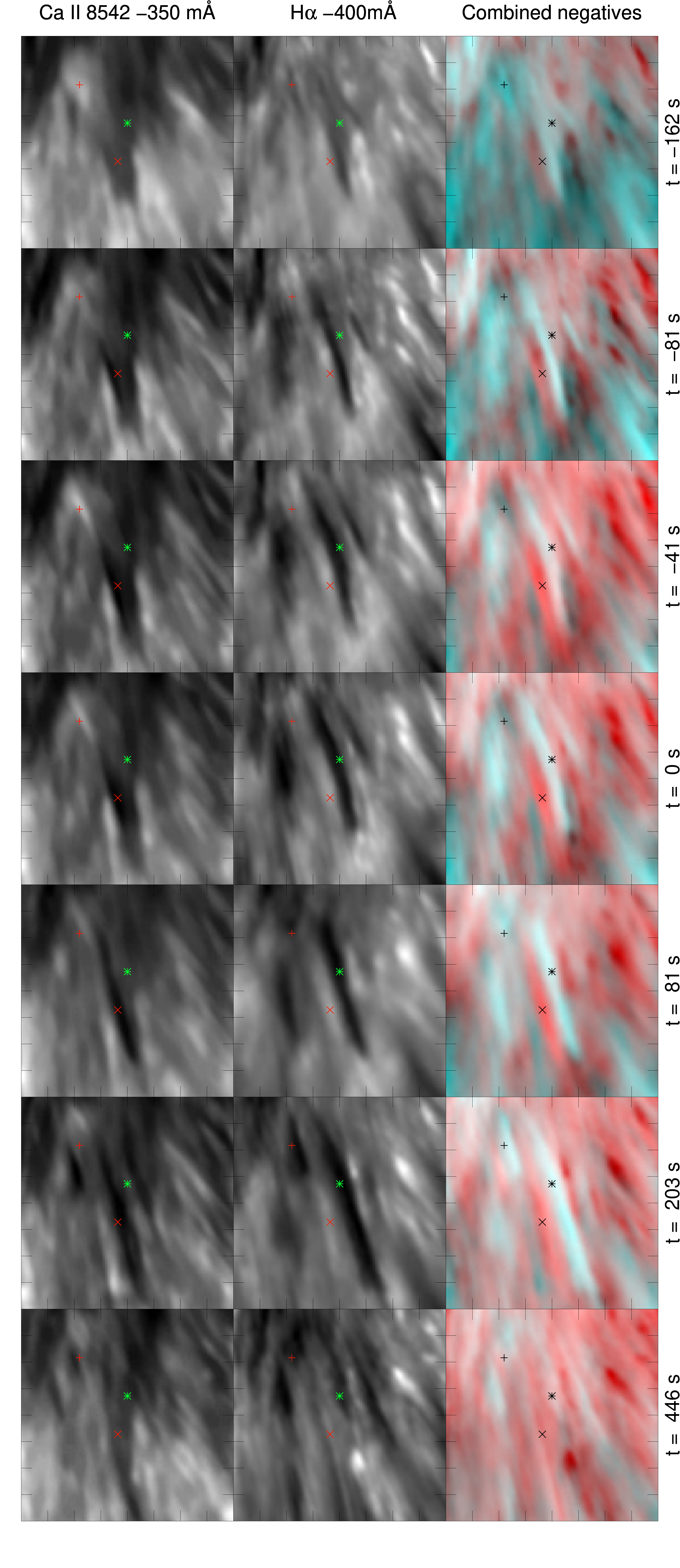} 
\caption{\label{fig:time_evolution_darkening_example_B21} 
PMJ B21 and its associated \efft\ and \halpha\ line dark features. The layout is identical to that of Fig. \ref{fig:time_evolution_darkening_example_A10}.}
\end{figure}

Figure \ref{fig:time_evolution_darkening_example_B21} shows the \efft\ and \halpha\ line dark feature evolution of PMJ B21, analogous to what is shown in Fig.  \ref{fig:time_evolution_darkening_example_A10}.

\begin{figure}[!ht]
\centering
\includegraphics[width=8.2cm]{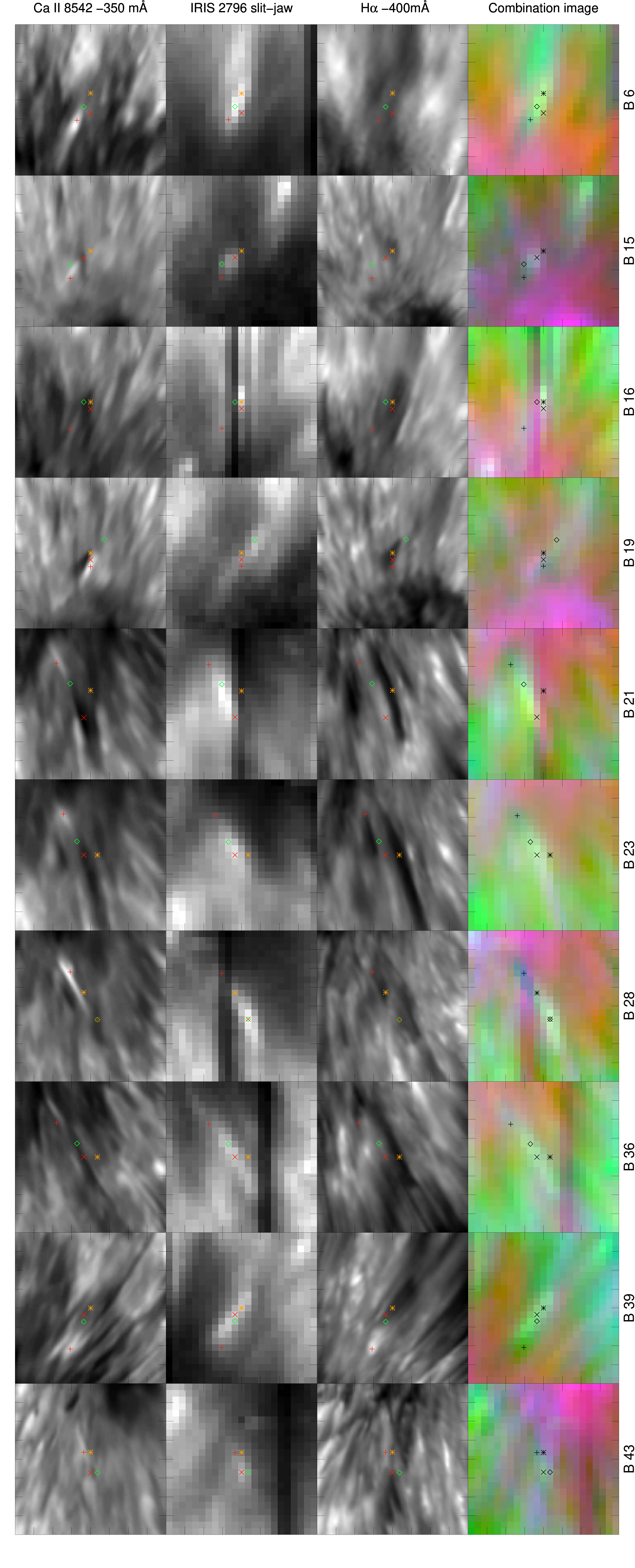} 
\caption{\label{fig:dark_examples_30th} 
PMJs drawn from dataset B for which dark features in the \efft\ blue wing (first column) and the \halpha\ inner blue wing (third column) overlap with brightening in the \mgii\ SJI channel (second column). The layout is identical to that of Fig. \ref{fig:dark_examples_29th}.}
\end{figure}

Figure \ref{fig:dark_examples_30th} shows example PMJs from dataset B with apparent \efft\ and \halpha\ line dark features that overlap with \mgii\ SJI channel bright features, analogous to those shown in Fig. \ref{fig:dark_examples_29th}. 

\begin{figure}[!th]
\centering
\includegraphics[width= 7.5cm]{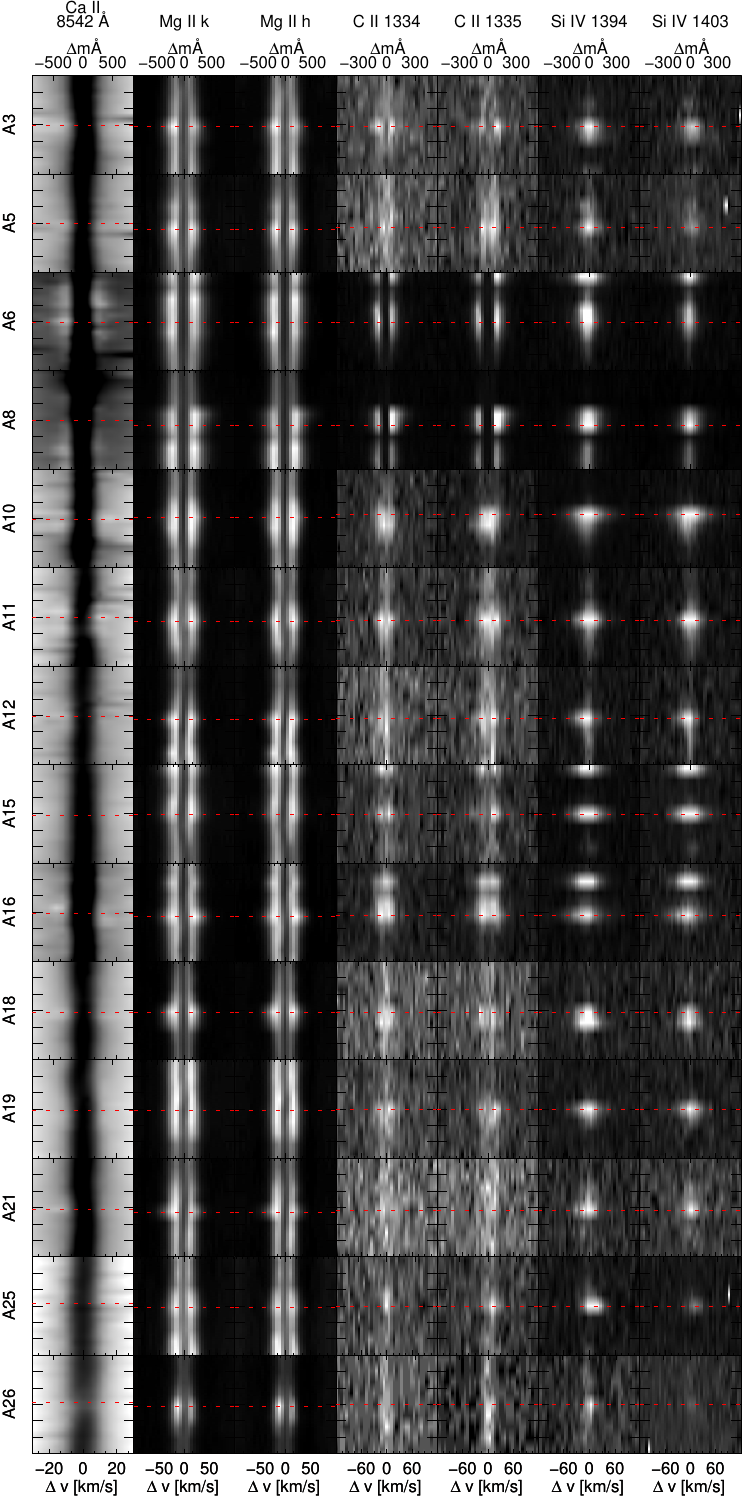} 
\caption{\label{fig:timeslices_29th} Spectral timeslices for a selection of PMJ events from dataset A that display clearly visible signals. The layout is identical to that of \ref{fig:timeslices_30th}.}
\end{figure}

Figure \ref{fig:timeslices_29th} shows timeslices for PMJs in dataset B for which intensity brightenings are apparent for most spectral lines, analogous to those shown in Fig. \ref{fig:timeslices_30th}.

%======================================= fig A1
%=======================================

%======================================= fig A2
%=======================================

%======================================= fig A3
%=======================================

\section{Information on online supplementary videos}\label{sec:appendix_videos}

The videos showing all PMJs at only peak brightness for dataset A and B are available in \href{https://www.mn.uio.no/astro/personer/vit/rouppe/movies/drews_2020_appa_video1.mp4}{Video 1} and 
\href{https://www.mn.uio.no/astro/personer/vit/rouppe/movies/drews_2020_appa_video2.mp4}{Video 2,}
respectively.

The videos detailing the temporal evolution of PMJs in dataset A and B are available in \href{https://www.mn.uio.no/astro/personer/vit/rouppe/movies/drews_2020_appa_video3.mp4}{Video 3} and 
\href{https://www.mn.uio.no/astro/personer/vit/rouppe/movies/drews_2020_appa_video4.mp4}{Video 4,}
respectively. 

Various plots are provided in the videos; all studied spectral lines (except the \halpha\ line), time- and space-slices in these lines, bisector plots, \mgii\ h \& k Dopplergrams, RGB images of the given PMJs that are marked with spectral profile sampling positions, and \efft\ blue inner line wing images of the PMJs.

Videos 1 and 2 show each PMJ for only the peak brightness time frame.
Videos 3 and 4 show the temporal evolution of all PMJs in each dataset; each PMJ is shown for 11 time frames, corresponding to five IRIS SJI time indexes before and after the nominal peak brightness time index, as well as the peak brightness time index of the given PMJ itself. This corresponds to showing the diagnostics corresponding to approximately 3 min 36 s for each PMJ. PMJs that appear close to the start and end times of our observation times were truncated in time accordingly.

\begin{sidewaysfigure*}[!ht]
\centering
\includegraphics[height=13.0cm]{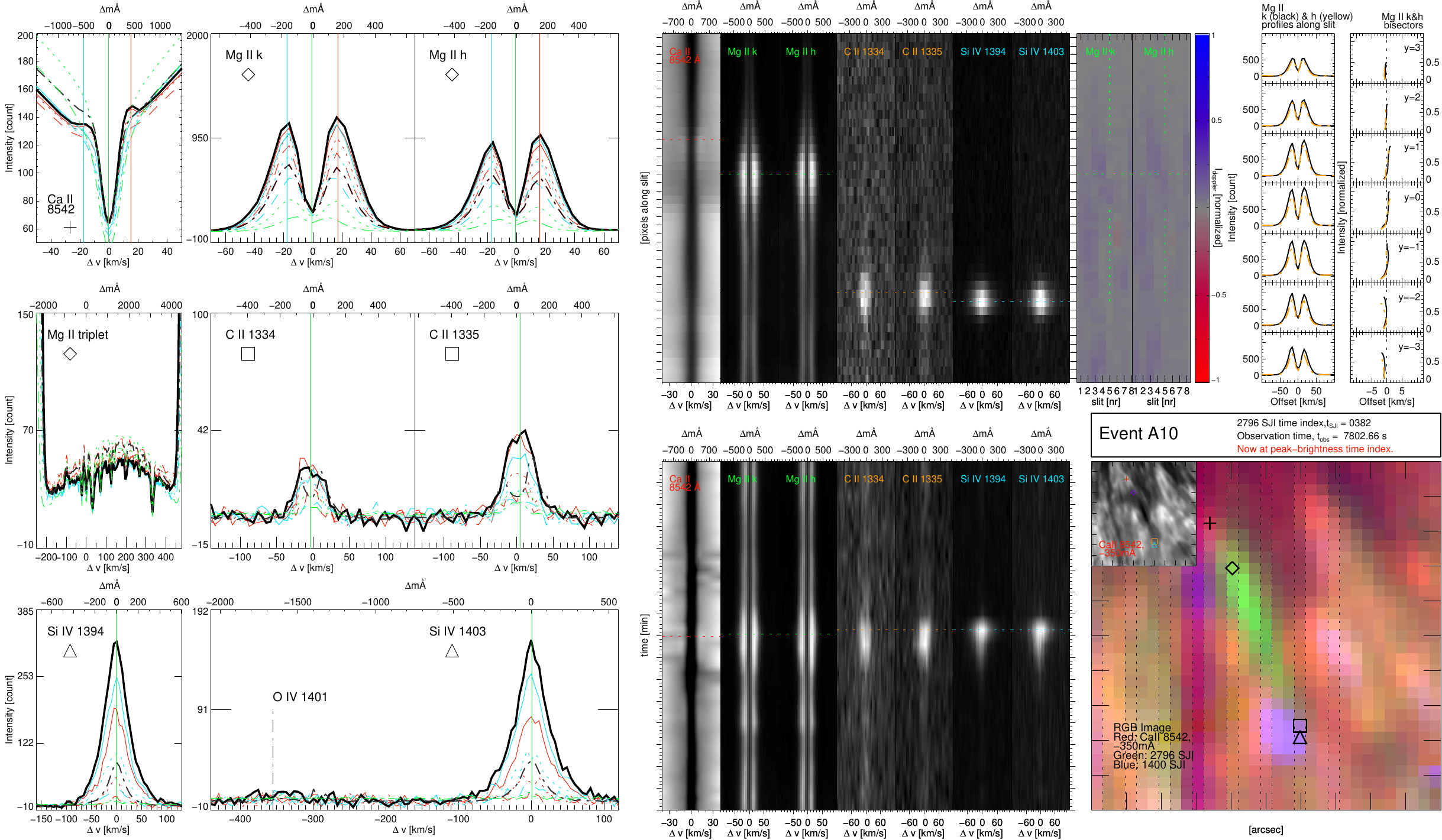} 
\caption{\label{fig:vid_example} Peak-intensity time frame for PMJ A10 from the temporal evolution video for individual PMJs in dataset A. The layout is identical for datasets A and B. The spectral line profiles are shown on the left-hand side; panels left to right and top to bottom show profiles for the \efft\ line, the \mgii\ k and h lines, the \mgii\ triplet, the \cii\ 1334 \AA\ and 1335 \AA\ lines, and the \siiv\ 1394 \AA\ and 1403 \AA\ lines (the latter has \oiv\ indicated). The different linestyles denote the following: black solid line, the PMJ line profile at the sampling position; blue and red profiles were sampled at positive and negative vertical offsets from the main sampling position, respectively, with a different line style for each one-pixel difference; solid lines (one-pixel offset); dotted lines  (two-pixel offset); dashed line (three-pixel offset); black-dashed line, the average PMJ line profile for the dataset; green-dotted line, the average line profile across all pixels in the observations; and green-dash-dotted line, the average penumbral line profile. Positions of line cores (vertical solid green lines), red peaks (vertical solid red lines), or blue peaks (vertical solid blue lines) are marked for those spectral lines where applicable.
Shown to the immediate right of the spectral profiles are space-slices (top) and time-slices (bottom) of the studied spectral lines; each panel is marked accordingly. Horizontal lines mark the sampled y-position for the space-slice panels and the sampling time for the time-slice panels. Given to the right of the space-slices are Dopplergrams in the \mgii\ k \& h lines, each panel is marked accordingly, with vertical and horizontal lines surrounding the \mgii\ PMJ sampling position and the three pixels below and above it. To the right of the Dopplergrams are individual \mgii\ h (yellow-dashed line) and k (black solid line) spectral profiles for seven pixels along the PMJ IRIS raster slit, with the corresponding bisectors for each spectral line, which are given to their right for both lines. Bisectors that have mean values > $|\pm 1|$ \kms\ are marked with a blue (negative value) or red circle (positive value). 
To the right of the timeslices we give the current RGB image centred on the PMJ; the sampling positions for each spectral line is indicated with symbols as given in the spectral line panels. Vertical dashed lines indicate the IRIS raster slit positions. An insert in the upper left shows the same FOV and sampling positions, but in the inner blue wing of the SST \efft\ line. Above the RGB image, the PMJ identification and the current sampling IRIS slitjaw image index as well as the absolute observation time in seconds are given.
}
\end{sidewaysfigure*}

In Fig. \ref{fig:vid_example} we show a frame drawn from the video for dataset A. It shows plots and images for the time of peak brightness for PMJ A10. The caption gives details for all plots, and the layout is valid for both the peak-brightness-only videos (1 \& 2) and for all time frames in both PMJ temporal evolution videos (3 \& 4) for both datasets A and B.

%======================================= fig A4
%=======================================

\end{appendix}

\end{document}